%% file: main.tex
\documentclass[10pt,conference]{IEEEtran}

\usepackage{booktabs}
\usepackage{microtype}
\usepackage{xspace}
\usepackage{listings}
\usepackage{amsmath}
\usepackage{graphicx}
\usepackage{multirow}
\usepackage{xcolor}
\usepackage{cite}
\usepackage{url}
\usepackage{hyperref}
\usepackage{cleveref}

\makeatletter
\long\def\@makecaption#1#2{%
\ifx\@captype\@IEEEtablestring%
\footnotesize\bgroup\par\centering\@IEEEtabletopskipstrut{\normalfont\footnotesize #1}\\{\normalfont\footnotesize #2}\par\addvspace{0.5\baselineskip}\egroup%
\@IEEEtablecaptionsepspace
\else
\@IEEEfigurecaptionsepspace
\setbox\@tempboxa\hbox{\normalfont\footnotesize {#1.}\nobreakspace\nobreakspace #2}%
\ifdim \wd\@tempboxa >\hsize%
\setbox\@tempboxa\hbox{\normalfont\footnotesize {#1.}\nobreakspace\nobreakspace}%
\parbox[t]{\hsize}{\normalfont\footnotesize\noindent\unhbox\@tempboxa#2}%
\else%
\ifCLASSOPTIONconference \hbox to\hsize{\normalfont\footnotesize\hfil\box\@tempboxa\hfil}%
\else \hbox to\hsize{\normalfont\footnotesize\box\@tempboxa\hfil}%
\fi\fi\fi}
\makeatother

\newcommand{\etal}{et~al\mbox{.}\xspace}

\newif\ifsupplementary
\supplementarytrue
\newcommand{\appref}[1]{\ifsupplementary\cref{#1}\else the supplementary material\fi}

\begin{document}

\title{Correct but Slow: An Empirical Study of the {GPU} Kernel Evaluation Gap in Modern Domain-Specific Languages}

\author{\IEEEauthorblockN{Tingxi Li, Ravishka Rathnasuriya, and Wei Yang}
\IEEEauthorblockA{\textit{The University of Texas at Dallas}\\
Richardson, TX, USA\\
\{tingxi.li, ravishka.rathnasuriya, wei.yang\}@utdallas.edu}}

\hypersetup{
  pdftitle={Correct but Slow: An Empirical Study of the GPU Kernel Evaluation Gap in Modern Domain-Specific Languages},
  pdfauthor={Tingxi Li, Ravishka Rathnasuriya, Wei Yang}
}

\maketitle
\pagestyle{plain}

\begin{abstract}
  \input{tex/abstract}
\end{abstract}


\section{Introduction}
\label{sec:introduction}
\input{tex/introduction}

\section{Background}
\label{sec:background}
\input{tex/background}

\section{Methodology}
\label{sec:methodology}
Studying what correctness-gated evaluation misses requires three ingredients: kernels that pass the gate, strong library baselines to measure them against, and instrumentation that attributes any gap to a cause. We therefore construct a kernel suite covering the main computation patterns in modern models, compare DSL implementations against library-backed PyTorch baselines under matched configurations, and profile both end-to-end latency and low-level hardware behavior.

\input{tex/methodology}

\section{The Evaluation Gap (RQ1)}
\label{sec:evaluation}
\input{tex/evaluation}

\section{Root Causes of the Hidden Gap (RQ2)}
\label{sec:analysis}
\input{tex/analysis}

\section{Guidance Without a Comprehensive Benchmark (RQ3)}
\label{sec:mitigation}
\input{tex/mitigation}

\section{Discussion}
\label{sec:discussion}

Our results recast DSL kernel performance as an evaluation problem: the largest gaps hide behind correctness-only gates, most are repairable authoring artifacts, and the remainder marks where compilers and libraries genuinely differ. We draw out implications for DSL and compiler developers, practitioners, and benchmark designers, then state the study's limitations and threats to validity.
\input{tex/discussion}

\section{Related Work}
\label{sec:related}
\input{tex/related_work}

\section{Conclusion}
\label{sec:conclusion}
\input{tex/conclusion}

\clearpage
\bibliographystyle{IEEEtran}
\bibliography{references}

\clearpage
\appendices
\crefalias{section}{appendix}
\crefalias{subsection}{appendix}
\input{tex/appendix}

\end{document}

%% file: tex/abstract.tex
Modern GPU domain-specific languages (DSLs), such as Triton and TileLang, are increasingly used to implement specialized deep-learning kernels and as target languages for automated kernel-generation systems. Existing DSL-kernel evaluations establish correctness through reference-based numerical validation---necessary, but silent on replacement quality: a functionally valid kernel may still fall far below the throughput of the optimized library operator it is intended to replace.

We study this correctness–performance gap using 22 Triton and TileLang kernels from five operator categories on NVIDIA A100 and GH200 GPUs, asking whether correctness-based evaluation identifies kernels unsuitable as library replacements, why such failures occur, and how they can be detected without exhaustive benchmark coverage. The study yields three results. \emph{First}, correctness-based evaluation can admit severe slowdowns: an idiomatic TileLang LayerNorm kernel passes KernelBench’s correctness check while running more than 300× slower than the PyTorch baseline. \emph{Second}, the causes differ by kernel family. TileLang normalization and reduction slowdowns are mainly repairable authoring defects, such as sequential reductions and unnecessary dtype conversions, whereas convolution and large general matrix multiplication (GEMM) retain residual gaps after optimization due to code-generation and autotuning-coverage limits; vendor-library algorithm selection contributes only marginally. \emph{Third}, two lightweight checks---library-relative efficiency and roofline utilization---are complementary screening criteria: together they flag every functionally valid but inefficient kernel in our suite and separate repairable authoring defects from structural residuals.

%% file: tex/introduction.tex

Deep learning systems depend on GPU kernels that dominate execution time~\cite{acceleration_survey, optimization_survey}. Frameworks have historically delegated these kernels to vendor libraries such as cuBLAS and cuDNN~\cite{cublas2024, chetlur2014cudnn}, but modern architectures increasingly require fused, specialized, model-specific computations outside what those libraries expose~\cite{kernelfusion}. Domain-specific languages (DSLs) such as Triton~\cite{openai2021triton} and TileLang~\cite{wang2025tilelang} have emerged as the standard answer: they express kernels at the tile level and delegate memory movement, scheduling, and code generation to a compiler. Triton is already the lowering target for \texttt{torch.compile}~\cite{li2023torchcompile}, and DSLs are increasingly the output of large language model (LLM)-based kernel generators~\cite{kernelbench}---so whether DSL kernels can match vendor-library performance is now central to deploying them.

A DSL kernel measured against the vendor library may be faster, comparable, or slower by orders of magnitude, yet the measurement alone does not reveal the cause: poor authoring, a hardware ceiling already reached, or limits in the compiler's code generation or library maturity~\cite{williams2009roofline}. No systematic method separates these cases, and for the fused, model-specific operators that motivate DSLs in the first place, no strong vendor baseline exists to anchor the comparison at all~\cite{dao2023flashattention2}; closing this attribution gap requires tracing performance to its cause, not just observing it.

Existing benchmarks do not close this gap: they admit performance-poor kernels, and their coverage across DSLs, data types, shapes, and GPUs is too narrow to certify kernel quality. KernelBench~\cite{kernelbench} and TritonBench~\cite{li2025tritonbench}, two prevailing benchmarks, report performance as an outcome but gate only on correctness. A naive but idiomatic TileLang kernel passes KernelBench's correctness gate while running more than 300$\times$ slower than PyTorch; KernelBench's only reward-hacking guard never fires, because it watches for suspiciously \emph{fast} kernels. Both benchmarks also cover only one DSL, data type, shape, and GPU per task, and building one that spans the full space is not practical.

We address this evaluation gap in three steps.
First, we run KernelBench~\cite{kernelbench} end-to-end and analyze TritonBench's~\cite{li2025tritonbench} acceptance criteria, showing that both admit performance-poor kernels: the correctness gate accepts kernels orders of magnitude slower than the baseline, and coverage is too narrow to certify quality beyond the tested configuration.
Second, we study 22 kernels across five operator classes on an NVIDIA GH200 and an A100-SXM4-40GB (hereafter A100), using hardware counters to trace each performance gap to a specific cause: authoring, code generation, or library immaturity.
The dominant TileLang normalization gap turns out to be an authoring artifact, removable by correcting a single reduction idiom (a two-line \texttt{T.reduce} rewrite plus native-dtype I/O), while the convolution and large general-matrix-multiplication (GEMM) gaps are structural.
Third, building on these findings, we propose two lightweight heuristics---a library-comparability screen and a baseline-independent roofline anchor---together with a small set of recurring optimization patterns; they flag poor kernels and recover most of the gap without a comprehensive benchmark.

\input{tex/RQ_summary}

\smallskip\noindent This paper makes the following contributions:
\begin{itemize}
  \item \textbf{Evaluation gap (RQ1).} Prevailing DSL kernel benchmarks gate on correctness alone and admit kernels that run up to $300\times$ slower than the library baseline.
  \item \textbf{Hidden gap and causes (RQ2).} We trace the performance gap these benchmarks miss to specific authoring, code-generation, and library-maturity causes, each with a distinct hardware-counter signature.
  \item \textbf{Benchmark-free guidance (RQ3).} We give two lightweight evaluation heuristics and a set of recurring optimization patterns that flag poor kernels and guide their repair without a comprehensive benchmark.
\end{itemize}
\noindent Artifacts  are available at \url{https://github.com/tingxi-li/DSLPerfGap/tree/master} to support reproducibility.

%% file: tex/RQ_summary.tex

We organize the study around three research questions.

\smallskip\noindent\textbf{RQ1 (Evaluation gap).} Do existing DSL and LLM-generated-kernel
benchmarks distinguish efficient kernels from performance-poor ones, and how large is the
performance gap they admit for kernels that pass?

\smallskip\noindent\textbf{RQ2 (Root causes).} Which authoring, code-generation, and
library-maturity factors cause the hidden gap that RQ1 exposes?

\smallskip\noindent\textbf{RQ3 (Guidance without a comprehensive benchmark).} Can lightweight
heuristics---a library-comparability screen and a roofline anchor---plus recurring
optimization patterns reliably flag poor kernels and guide their repair?

%% file: tex/background.tex
\subsection{Tile-Based GPU DSLs}
\label{sec:bg:dsls}
A GPU executes thousands of threads grouped into \emph{thread blocks}, each resident on a \emph{streaming multiprocessor} (SM) with fast programmer-managed \emph{shared memory}; high-performance kernels \emph{tile} data~\cite{wolfe1989more} to reuse that scratchpad and stay compute-bound under the roofline model~\cite{williams2009roofline}.
Two Python-embedded DSLs raise this tiling abstraction above hand-written CUDA, and our study targets both.
Triton~\cite{tillet2019triton} makes the \emph{tile}---a statically shaped, multi-dimensional array operated on by all threads in a program instance---the unit of computation.
The Triton compiler, built on MLIR~\cite{lattner2021mlir}, performs memory coalescing, shared-memory allocation, thread swizzling, and software pipelining automatically while targeting PTX (NVIDIA's virtual assembly), and programmers tune tile configurations through \texttt{@triton.autotune}.
Triton has been the default lowering target for \texttt{torch.compile} since PyTorch 2.0~\cite{li2023torchcompile}.
Triton's convolution support is less mature: an im2col-style lowering (reshaping convolution windows into matrix columns so convolution runs as GEMM) exists but falls substantially short of cuDNN on common workloads~\cite{triton2022conv591}.
TileLang~\cite{wang2025tilelang} separates the \emph{algorithm} (what to compute) from the \emph{schedule} (how to map computation onto GPU resources) more explicitly than Triton, letting the programmer tune memory staging, thread layout, and pipeline depth without changing the functional description.
It compiles through Apache TVM with hardware-intrinsic support on NVIDIA and AMD; its convolution performance relative to cuDNN has not been systematically evaluated prior to this work.
Throughout, we compare both DSLs against the vendor libraries they aim to displace: cuBLAS~\cite{cublas2024} for dense GEMM and cuDNN~\cite{chetlur2014cudnn} for convolution and other deep-learning primitives, each dispatching its fastest implementation per shape via a runtime selector~\cite{cutlass2023}.

\subsection{Benchmarking DSL and LLM-Generated Kernels}
\label{sec:bg:tritonbench}
Custom GPU kernels are increasingly produced not by hand but by automated tools: \texttt{torch.compile} lowers operators to Triton~\cite{li2023torchcompile}, and a growing body of research uses large language models (LLMs) to generate kernels directly~\cite{kernelbench,li2025tritonforge}.
This has made kernel benchmarks the de facto arbiters of kernel quality, and two benchmarks dominate the DSL and LLM-generated setting.
TritonBench~\cite{li2025tritonbench} curates production-grade Triton kernels and reports correctness plus a roofline-anchored GPU-efficiency metric; KernelBench~\cite{kernelbench} accepts a kernel that reproduces a PyTorch reference on randomized inputs.
What each benchmark measures, what it gates on, and what its gate admits are the subject of \cref{sec:evaluation}.

%% file: tex/methodology.tex
\subsection{Kernel Suite}
\label{sec:meth:kernels}

Our benchmark suite covers five operator categories that capture the dominant computation patterns in modern deep learning: matrix multiplication, attention, convolution, normalization, and element-wise or reduction operators.
In total, the suite contains 22 kernels: three \textbf{GEMM} workloads (dense matmul, batched matmul, fused linear+activation), one \textbf{attention} workload (scaled dot-product / FlashAttention variants), one \textbf{convolution} workload (Conv2d, $1\times1$--$7\times7$ filters including depthwise and strided), two \textbf{normalization} workloads (LayerNorm and RMSNorm), and fifteen \textbf{element-wise or reduction} workloads.
Triton kernels are drawn from TritonBench~\cite{li2025tritonbench} (curated from its GitHub channel of 184 kernels across 95 repositories) and, for LayerNorm, the TorchInductor reference; all TileLang suite kernels are our re-implementations of the same operators, following the interfaces of the TritonBench versions and the idioms of the TileLang example repository~\cite{wang2025tilelang}.
Per-kernel provenance and full selection criteria are in \appref{app:repro:kernels}; the authorship threat this creates is discussed in \cref{sec:disc:limitations}.

\subsection{Baseline Construction}
\label{sec:meth:baselines}

For each workload, we compare DSL implementations against the strongest available vendor-library path through PyTorch; per-category baseline specifications are in \appref{app:repro:baselines}.
For attention, we use PyTorch scaled dot-product attention with the FlashAttention backend; because the Triton and TileLang implementations are not algorithmically identical to this baseline, we report attention results separately.

\subsection{Profiling Setup}
\label{sec:meth:profiling}

We measure both end-to-end latency (which drives all library comparisons) and hardware performance counters (which support root-cause analysis, RQ2).

\paragraph{End-to-end throughput}
We measure host-synchronized GPU execution time. Locked-clock runs pin graphics and memory clocks to fixed frequencies with run-to-run relative standard deviation $\leq 0.9\%$; each reported table states whether its measurements are clock-locked (protocol and clock settings in \appref{app:repro:meas}).

\paragraph{Hardware performance counters}
For root-cause analysis, we collect Nsight Compute~\cite{nsightcompute2024} (NCU) profiles grounded in seven counters (definitions in \appref{app:repro:meas}), each from a single execution with stability verified within 2\% across five runs.

\paragraph{Correctness}
Before any timing, we validate every DSL kernel against its library baseline at per-dtype tolerances (\texttt{fp32}~$10^{-5}$, \texttt{fp16}~$10^{-3}$, \texttt{bf16}~$10^{-2}$, relaxed $2\times$ for reductions) and an edge-case suite (NaN/Inf/denormal/all-equal) across all 22 kernels with 0 crashes.
FP32 TileLang GEMM is excluded from timing: \texttt{T.gemm} silently lowers float32 to TF32 on SM80, failing an exact $10^{-5}$ check on 31.8\% of outputs while a manual FMA kernel passes (a precision-lowering artifact, reproducible via \texttt{exp\_fp32\_gemm.py} in the artifact), so all TileLang GEMM is measured in FP16.

\subsection{RQ3 Optimization Methodology}
\label{sec:meth:optimization}

To produce the optimized kernels evaluated in \cref{sec:mitigation}, we use AKO~\cite{ako2026}, an existing LLM-based agentic kernel-optimization harness, as an off-the-shelf tool under a fixed protocol (correctness-gated iterations, large-input shapes, no language switching) to instantiate the root-cause-derived optimization patterns of \cref{sec:analysis}.
Optimized kernels are validated provenance-agnostically against the same per-dtype tolerances and Nsight Compute counters, and recovered performance is reported as a lower bound on user-space-recoverable gain.

\subsection{Metrics}
\label{sec:meth:metrics}

Our primary metric is \textbf{library efficiency},
\[
E_{\mathrm{lib}} = \frac{t_{\mathrm{lib}}}{t_{\mathrm{DSL}}} \times 100\%,
\]
where $t_{\mathrm{lib}}$ is the latency of the cuBLAS or cuDNN baseline and $t_{\mathrm{DSL}}$ that of the corresponding DSL implementation under the same hardware and input configuration; 100\% indicates parity, lower values a slower DSL kernel.
Secondary metrics (throughput, hardware efficiency, memory bandwidth utilization) are used qualitatively in root-cause analysis (RQ2) and are not tabulated separately.

\subsection{Evaluation-Gap and Heuristic Measurement}
\label{sec:meth:evalgap}

To test whether existing benchmarks admit slow kernels (RQ1), we run each naive DSL kernel through KernelBench's evaluator, which overrides the candidate's input generators with the reference's and checks output equivalence on randomized inputs; we record the pass/fail verdict and measured slowdown against the PyTorch baseline (harness details in \appref{app:repro:meas}).
To anchor the RQ3 heuristic in a baseline-independent signal, we model the bytes moved for memory-bound kernels or the floating-point operations (FLOPs) for compute-bound kernels, then divide the achieved work rate by the relevant hardware peak (HBM bandwidth or FP16 Tensor-Core throughput) to obtain its roofline fraction.

\subsection{Experimental Setup}
\label{sec:meth:setup}

\paragraph{Hardware}
We measure on two primary architectures: the A100-SXM4-40GB (Ampere, \texttt{sm\_80}), used for suite magnitude (\cref{sec:evaluation}) and root-cause counters (\cref{sec:analysis}), and the GH200 (Grace Hopper, \texttt{sm\_90}), used for the evaluation-gap demonstration (\cref{sec:eval:passesslow}) and roofline heuristic (\cref{tab:roofline}).
Full specifications and clock-lock settings are in \appref{app:repro:stack}.

\paragraph{Software}
We use PyTorch~2.8.0+cu128, Triton~3.4.0, TileLang~0.1.6.post1, bundled cuDNN, and Nsight Compute~2026.2.0; the complete version list is in \appref{app:repro:stack}.

%% file: tex/evaluation.tex
This section answers RQ1: \emph{do existing benchmarks distinguish efficient DSL kernels from performance-poor ones, and along which dimensions do they fall short?}
We survey what the prevailing benchmarks measure (\Cref{sec:eval:survey}), show that a correct-but-slow kernel passes their gate unflagged (\Cref{sec:eval:passesslow}), and quantify the hidden gap across the 22-kernel suite (\Cref{sec:eval:overview} onward).

\subsection{What Existing Benchmarks Measure}
\label{sec:eval:survey}

The two benchmarks that dominate DSL and LLM-generated GPU-kernel evaluation are KernelBench~\cite{kernelbench} and TritonBench~\cite{li2025tritonbench}, and both treat performance as a \emph{reported outcome} rather than an \emph{acceptance criterion}.
KernelBench, the standard LLM kernel-generation target, accepts a candidate on \emph{correctness alone}: measured speedup is recorded---and aggregated in the performance-aware \texttt{fast\_p} leaderboard metric---but a slow kernel is never rejected, so our critique targets the \emph{gate}, not the metric: every passing kernel enters the accepted pool that downstream users and kernel-generation loops treat as usable.
TritonBench reports a roofline-anchored efficiency metric---a precedent \Cref{sec:mitigation} builds on---but covers only Triton, one data type and shape per operator, and a single GPU architecture.
Neither benchmark spans the full (DSL $\times$ data type $\times$ shape $\times$ architecture) space, and extending one would be costly: re-tuning a single kernel at one deployment shape took 211~s versus under a second at a small shape, a cost that multiplies across every cell.
A passing kernel certifies correctness, not performance quality.

\subsection{Passing Is Not Fast}
\label{sec:eval:passesslow}

We pass idiomatic DSL kernels through the KernelBench correctness gate and time them.
On the GH200, TileLang LayerNorm passes the correctness gate on all five shapes but runs $323\times$ slower than PyTorch at the large shape under locked clocks ($51.5$~ms vs.\ $0.16$~ms); TileLang RMSNorm passes and is $117\times$ slower. A deliberately constructed worst-case LayerNorm variant reaches $1293\times$ ($176.4$~ms, unlocked); even a naive argmax passes and is $16.7\times$ slower.\footnote{The LayerNorm slowdowns quoted in this paper differ only in GPU, kernel variant, and comparison target---A100 vs.\ PyTorch: $299\times$ (tuned idiomatic, 97.0~ms); A100 naive$\to$optimized: $380\times$ (tuned) and $1347\times$ (untuned, 364$\to$0.270~ms, locked; \cref{tab:mitigation}); GH200 vs.\ PyTorch: $323\times$ (locked), $329\times$ (suite profile), $1293\times$ (worst-case variant, unlocked); GH200 naive$\to$optimized: $1541\times$ (worst-case; \cref{tab:roofline}).}
In every case the gate returns \texttt{correct=True} and the more-than-$10\times$ reward-hacking guard never triggers, because the kernels are slow, not fast.
These are not adversarial constructions: they differ from a competent implementation in a single reduction primitive (\Cref{sec:analysis:jit}); \Cref{fig:serial-vs-reduce} shows the two-line fix. The A100 suite exhibits the same pattern; per-architecture magnitudes are reported throughout.

\subsection{The Hidden Gap These Benchmarks Admit}
\label{sec:eval:overview}

Every kernel in the 22-kernel suite passes the same correctness gate (\cref{sec:meth:profiling}), so each sub-parity efficiency below is invisible to a correctness-only evaluator; root-cause labels (RC0--RC4, defined in \cref{sec:analysis}) annotate each finding so RQ1 magnitudes can be read against RQ2 causes.

\begin{figure}[t]
  \centering
  \IfFileExists{figures/overview_efficiency.pdf}{%
    \includegraphics[width=.75\linewidth]{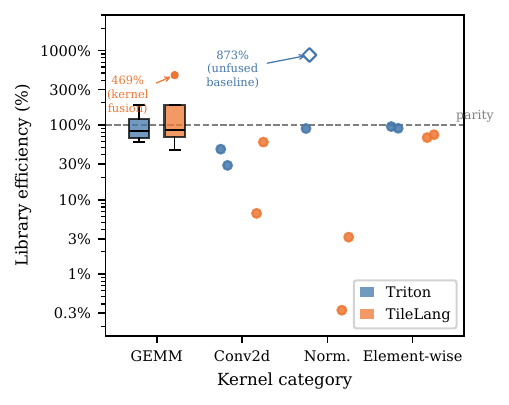}%
  }{%
    \fbox{\parbox[c][6cm][c]{\linewidth}{\centering
      \textit{[Figure: Library efficiency (\%) by kernel category and DSL.
      Violin or box plot. To be generated from profiling data.]}}}%
  }
  \caption{Library efficiency (\%) of Triton and TileLang kernels vs.\ cuBLAS/cuDNN
    on the A100, by category (log scale; dashed line = parity; tuned suite profile as \Cref{tab:summary}).
    GEMM shows IQR boxes; other categories per-kernel dots.
    $\diamond$~873.0\% (Triton, Norm.) is \texttt{rms\_norm}, an unfair-baseline artifact; 468.5\% (TileLang, GEMM) is \texttt{fused\_linear\_activation}, a genuine fusion advantage (\Cref{sec:eval:gemm}).
    Attention excluded: the Triton (tiled FlashAttention-2~\cite{dao2023flashattention2}) and TileLang (untiled $O(n^2)$) kernels are algorithmically different.}
  \label{fig:overview}
\end{figure}

\Cref{fig:overview} summarizes library efficiency (\%) for all kernels in the suite, broken down by DSL and kernel category (tuned suite profile, as \cref{tab:summary}).
The key finding is that the performance gap is not uniform: Triton is broadly competitive for element-wise and normalization kernels (LayerNorm 90.2\%, element-wise $\sim$69--96\%; the 873.0\% RMSNorm outlier is a fusion artifact) but trails on convolution (28.9\%) and, more mildly, on large square GEMM (59.7\%); TileLang is competitive across GEMM, convolution, and element-wise shapes but collapses on normalization and the softmax and index-returning reductions (LayerNorm 0.33\%, softmax 0.9\%, argmax 5.9\%).
RQ3 (\Cref{sec:mitigation}) shows that reduction and normalization gaps are authoring artifacts, whereas the convolution and large-GEMM gaps persist as genuine residuals.

\phantomsection\label{sec:eval:gemm}

\medskip\noindent\textbf{Finding: On the A100, Triton holds 59.7--97.9\% of cuBLAS on GEMM and TileLang 78.1--94.2\%; both DSLs are weakest at $16384^2$ and near parity on the batched shape (\Cref{tab:summary}, \Cref{fig:overview}).}

The extremes differ by shape rather than by DSL (\Cref{tab:summary}): both DSLs are weakest on the $16384^2$ square matmul (Triton 59.7\%, TileLang 78.1\%) and strongest on the $128\times2048^2$ batched shape (97.9\%/94.2\%), while a fused linear-plus-activation kernel pushes TileLang to $468.5\%$.
\Cref{sec:analysis} attributes Triton's $16384^2$ gap to missing autotune configurations (RC2a) compounded by absent cache blocking for the ${\approx}1.6$~GB working set, which leaves the kernel DRAM-bound where cuBLAS stays cache-resident (RC2b).
The fusion win reflects kernel design, not a compiler artifact---the single launch avoids PyTorch's intermediate allocation---and says nothing about shapes with no second operation to fuse.
The pattern holds on the GH200 (Triton 56--170\%, TileLang 60--163\%; \Cref{tab:summary}): batched and fused shapes reach or exceed parity while $16384^2$ stays the weakest shape on both architectures (Triton ${\sim}60\%$/$56\%$), so the large-square-GEMM residual is a general tuning and cache-blocking gap (RC2a, RC2b), not A100-specific.

\phantomsection\label{sec:eval:conv}

\medskip\noindent\textbf{Finding: Both DSLs trail cuDNN on convolution, but unevenly: Triton 13.3--47.4\% and TileLang 43.6--59.3\% across $1\times1$--$7\times7$ filters at the large shape. Both collapse to $\le$5.5\% on depthwise convolution, which lowers to 512 per-group GEMM launches---per-launch JIT overhead plus absent strided-access vectorization (RC1).}

\Cref{tab:summary} reports the convolution efficiencies, with Triton falling monotonically from 47.4\% ($1\times1$) to 13.3\% ($7\times7$) as the gap widens on the Winograd-ineligible larger filters (\Cref{sec:analysis}).
Nsight Compute shows the conv gap is occupancy- and coalescing-bound (RC1 + RC2), not a register-spill problem: \texttt{n\_spills = 0} at every filter size for both DSLs.
On the GH200, Triton convolution stays comparably weak (12.1\% at $3\times3$, $\le$9.8\% at the larger filters), while TileLang convolution is competitive on both architectures (A100 43.6--59.3\%, GH200 32.9--71.7\%; \Cref{tab:summary}).
The persistent residual is therefore the \emph{Triton} convolution gap, which holds on both architectures.

\phantomsection\label{sec:eval:norm}

\medskip\noindent\textbf{Finding: Triton normalization kernels match or substantially outperform PyTorch's eager paths; TileLang normalization collapses to 0.3--3.2\% of PyTorch throughput at the tested size ($8192 \times 8192$), with LayerNorm 299$\times$ slower.}

\Cref{tab:summary} reports normalization category medians.
On the A100, Triton \texttt{layer\_norm} reaches 90.2\% of PyTorch throughput at $8192^2$; TileLang \textbf{LayerNorm} runs $299\times$ slower ($97.0$~ms vs.\ $0.324$~ms in the tuned suite profile; the untuned kernel that \Cref{sec:mitigation} repairs is slower still at 364~ms under locked clocks).
The TileLang collapse persists on the GH200 ($329\times$ in the suite profile), confirmed within noise by clock-locked re-timing (\Cref{sec:eval:passesslow}).
\Cref{sec:analysis:jit} attributes this collapse to serialized memory-latency stalls under \texttt{T.serial} reduction loops.

\phantomsection\label{sec:eval:tuning}

\medskip\noindent\textbf{Finding: Hardware-agnostic heuristic tuning of the block-tile grid yields $\Delta\approx0$pp for both DSLs on every kernel---default and tuned latencies are near-identical, and the convolution gap is unchanged (28.9\%).}

Re-evaluating all 22 kernels with heuristically tuned configurations for both DSLs changed nothing ($\Delta = 0$pp on every row, convolution included): the gaps are insensitive to block-tile shape.
For convolution the limitation is structural code generation, which no tile shape repairs (RC1, RC3); for large GEMM the effective lever lies outside this grid entirely, in the scheduling dimensions that the expanded search below sweeps (RC2a).

The tension with the matmul speedup of \Cref{sec:mitig:other} is a search-space distinction, not a shape distinction: the 12-config grid varies only the block-$M$/$N$/$K$ tile dimensions, while the expanded search additionally sweeps \texttt{GROUP\_SIZE\_M} (the ``L2-swizzle'' of RC2a), \texttt{num\_warps}, and \texttt{num\_stages}.
That search recovers Triton from 59.7\% to 81.9\% at $16384^2$ and from 71.5\% to 77.3\% at $4096^2$; the optimal configuration is simply absent from the default grid (RC2a).

\phantomsection\label{sec:eval:summary}

\begin{table}[t]
  \centering
  \caption{Library efficiency (\%) per kernel category and DSL on the A100 and GH200
    (both unlocked tuned profile), computed over large-size benchmark configurations.
    Cells report the category median, or the min--max range where the category spans several kernels.
    Attention excluded from Overall (\Cref{fig:overview}).}
  \label{tab:summary}
  \resizebox{\dimexpr\columnwidth-4pt\relax}{!}{%
  \begin{tabular}{lcccc}
    \toprule
    & \multicolumn{2}{c}{A100-SXM4} & \multicolumn{2}{c}{GH200} \\
    \cmidrule(lr){2-3}\cmidrule(lr){4-5}
    Kernel category    & Triton & TileLang & Triton & TileLang \\
    \midrule
    GEMM               & 59.7--97.9\% & 78.1--94.2\% & 56--170\%\rlap{$^\P$} & 60--163\%\rlap{$^\P$} \\
    Attention          & \multicolumn{4}{c}{\textit{baselines non-equivalent}} \\
    Convolution        & 28.9\% & 59.3\% & 12.1\% & 54.1\% \\
    Normalization      & 90.2\%\rlap{$^\ddagger$} & 0.3--3.2\%\rlap{$^\ddagger$} & 74.8\% & 0.3--3.3\% \\
    Element-wise       & $\sim$69--96\%\rlap{$^\S$} & $\sim$67--74\% & $\sim$66--98\% & $\sim$62--68\% \\
    \midrule
    \textbf{Overall (excl.\ attn.)} & \textbf{$\sim$98\%} & \textbf{$\sim$68\%} & \textbf{$\sim$98\%} & \textbf{$\sim$67\%} \\
    \bottomrule
  \end{tabular}}%
  {\scriptsize
    $^\ddagger$ Triton's cell reports \texttt{layer\_norm} only (90.2\%); \texttt{rms\_norm}'s 873.0\% is excluded as an unfused-baseline artifact (\Cref{sec:eval:norm}). TileLang's 0.3--3.2\% spans both norm kernels.
    $^\P$ GH200 GEMM ranges include the fused linear+activation shape; the A100 fused outlier (468.5\%) is excluded from the range and plotted separately (\Cref{fig:overview}).
    $^\S$ Element-wise ranges cover the central cluster; below-range Triton outliers (\texttt{argmax} 25.5\%, \texttt{log\_softmax} 45.3\%, \texttt{index\_select} 59.2\%) are discussed in the text.
    Overall = per-DSL median across all non-attention kernels at the large shape.}
\end{table}

The most striking asymmetry is TileLang's normalization collapse: $299\times$ on large LayerNorm (compounding RC0 deficiencies, \Cref{sec:analysis:jit}), persisting across architectures ($329\times$ on the GH200).
The A100 and GH200 give a consistent picture (\Cref{tab:summary}): TileLang's overall median is $\sim$68\% on the A100 and $\sim$67\% on the GH200, and the two gaps that survive on \emph{both} architectures are the TileLang normalization collapse and the Triton convolution residual (28.9\% at $3\times3$ on the A100, $\le$13\% on the GH200), not a fixed per-DSL ranking. Competitiveness still cannot be read off one GPU at the \emph{per-kernel} level: \Cref{sec:mitigation} shows an optimized \texttt{logsumexp} kernel that is fast on the A100 but register-spills on the GH200.

Within the element-wise and reduction category, most kernels cluster near the per-DSL medians in \Cref{tab:summary}; the notable Triton outlier is the index-returning reduction \texttt{argmax} (25.5\%), while fusion or baseline-path cases (\texttt{leaky\_relu}, \texttt{cross\_entropy}, \texttt{matrix\_transpose}, \texttt{linear+act}) exceed 100\% and are not vendor-library speedups.
Per-kernel efficiencies for all 22 benchmarked kernels appear in \appref{tab:perkernel}.

%% file: tex/analysis.tex
This section answers RQ2: \emph{for the kernels that pass existing correctness benchmarks (\cref{sec:eval:survey}) yet lag the vendor library by the margins in \cref{sec:evaluation}, what causes the hidden gap?}
We separate the causes by where the fix belongs: user-space \emph{authoring} (RC0a), compiler \emph{code generation} (RC0b, RC1, RC2a, RC3), and \emph{library maturity}---the tuning and algorithm selection a mature vendor library provides (RC2b, RC4).

\phantomsection\label{sec:analysis:jit}
\paragraph{RC0: TileLang Normalization and Reduction Deficiencies}
RC0 has two mechanisms, separated by where the fix lives: RC0a is a user-space authoring choice; RC0b is a compiler-codegen deficiency.

\paragraph{RC0a: TileLang Reduction Authoring (\texttt{T.serial} $\to$ \texttt{T.reduce})}
The original TileLang normalization kernels reduce over the normalized dimension with \texttt{T.serial} loops, serializing the accumulation with no inter-thread parallelism; the parallel alternative is \texttt{T.reduce} (\Cref{fig:serial-vs-reduce}).

\begin{figure}[t]
\begin{lstlisting}[language=Python, basicstyle=\ttfamily\footnotesize,
  commentstyle=\color{gray}\itshape]
# Before: sequential accumulation (T.serial)
for j in T.serial(n):          # n iterations, no parallelism
    mean_val[0] = mean_val[0] + row[j]

# After: parallel butterfly AllReduce (T.reduce)
T.reduce(row, mean_val, "sum", dim=0, clear=True)
\end{lstlisting}
\caption{RC0a fix: \texttt{T.serial}$\to$\texttt{T.reduce} in TileLang LayerNorm
  (mean pass; variance identical). With native-dtype I/O this recovers $1347\times$
  on the A100 (locked clocks; \cref{sec:mitig:norm}).}
\label{fig:serial-vs-reduce}
\end{figure}
The cost is exposed memory latency, not barriers: \texttt{warp\_stall\_long\_scoreboard} drops from 59.56 cycles to ${\approx}0$ under the fix while \texttt{warp\_stall\_barrier} is already ${\approx}0$ in both kernels; the GH200 shows the same signature (barrier ${=}0$, long-scoreboard ${=}49.95$, plus a $45.1$/$34.4$~GB local-load/store spill, RC3).
The \texttt{T.reduce} rewrite removes this latency source; with the native-dtype I/O fix, LayerNorm drops from 364~ms to 0.270~ms ($1347\times$, \cref{sec:mitig:norm}).

\paragraph{RC0b: Absent Vectorized Loads (Compiler Codegen)}
TileLang's compiled normalization kernels also fetch 16-bit bfloat16 elements individually instead of issuing 128-bit \texttt{LDG.128} loads, flooding the memory controller with discrete transactions; NCU confirms the scalar granularity for LayerNorm (50\% bytes/sector at one sector per request, DRAM throughput 59.2\%).
Unlike RC0a, this deficiency requires a compiler-codegen fix.

\phantomsection\label{sec:analysis:vec}
\paragraph{RC1: Absent Vectorization of Strided Memory Accesses}
Ampere reaches peak bandwidth only via 128-bit vector loads (\texttt{LDG.128}); Hopper's Tensor Memory Accelerator similarly requires aligned, contiguous descriptors.
For GEMM, Triton emits vectorized loads because the layout is row-major and tiles are 16-byte aligned.

Convolution breaks this in two ways: each output gathers a $(K_H\times K_W)$ window strided across the innermost NCHW dimension (PyTorch's default), breaking \texttt{LDG.128} alignment; and for $K_H\times K_W>1$ Triton's alias analysis cannot prove successive spatial-loop iterations are aligned, so it conservatively emits scalar 32-bit loads.
The consequence is a cascade noted in the community~\cite{triton2022conv591}: un-vectorized loads prevent \texttt{cp.async} from firing, so the software pipeline cannot overlap data movement with computation and the kernel stalls on global-memory latency at every tile boundary.

NCU confirms the absent vectorization on the A100: Triton conv2d achieves a load efficiency of only 12.55\% bytes/sector at 16.4 sectors per request, versus the \texttt{cp.async}-staged matmul path.\footnote{Triton matmul and cuBLAS read 0\% on this load counter because their \texttt{cp.async}/\texttt{LDGSTS} staging bypasses the \texttt{LDG} instruction the counter tracks.}

\phantomsection\label{sec:analysis:autotune}
\paragraph{RC2a: Configurations Absent From the Default Grid}
Triton's auto-tuner sweeps a programmer-specified configuration list
\{(\texttt{BLOCK\_M}, \texttt{BLOCK\_N}, \texttt{BLOCK\_K}, \texttt{num\_stages}, \texttt{num\_warps})\};
the reference GEMM tutorial's list achieves near-cuBLAS performance on common shapes~\cite{tillet2019triton}
but omits the L2-swizzle and large-shape configurations needed at scale.

The $16384^2$ matmul is the clearest case (\cref{sec:eval:gemm}): 59.7\% of cuBLAS at a shape absent from standard tutorial lists versus 97.9\% on a well-populated batched GEMM, and the expanded scheduling-parameter search of \cref{sec:eval:tuning} recovers it to 81.9\%---the missing configurations, not an exhausted budget, bound default performance.

\phantomsection\label{sec:analysis:l2cache}
\paragraph{RC2b: L2 Cache Residency}
For the $16384^2$ matmul, the combined A, B, C footprint (${\approx}1.6$~GB) vastly exceeds the A100's 40~MB L2 cache; cuBLAS employs hardware-specific cache-blocking (via \texttt{cuBLASLt}'s plan selector) while Triton's generic \texttt{tl.dot} compilation lacks equivalent heuristics, producing the ${\approx}1.7\times$ latency penalty observed ($33.5$~ms vs.\ $56.1$~ms).
NCU confirms the residency divide: Triton reaches only a 49.4\% L2 hit rate (DRAM-bound, 85.9\% throughput) whereas cuBLAS holds 80.5\% at 28.8\%.

Convolution adds filter-size degrees of freedom that default lists do not cover for $5\times5$/$7\times7$, and TileLang's ahead-of-time \texttt{num\_stages} cannot adapt to the register pressure large filters create.

\phantomsection\label{sec:analysis:regpressure}
\paragraph{RC3: Register Spill Collapses Occupancy (TileLang LayerNorm, Not Convolution)}
Each SM on the A100 exposes a 65{,}536-register file: a 128-thread block at 64 registers per thread fills it, preventing a second resident block and eliminating pipeline interleaving.
A natural hypothesis is that large-filter convolution hits this wall; NCU refutes it: Triton conv2d reports \texttt{n\_spills}\,=\,0 at every tested filter ($1\times1$ through $7\times7$), even though register usage rises with $K^2$ (up to 224 regs/thread at large shapes).
The spill is instead TileLang-LayerNorm-specific: the large LayerNorm kernel uses 254 registers per thread, drops to 12.4\% achieved occupancy, and spills 51.5~GB of local-load plus 34.4~GB of local-store traffic (A100; the GH200 mirrors it at 45.1/34.4~GB).
This register pressure---not any convolution effect---is what collapses occupancy below the level required for latency hiding.

\phantomsection\label{sec:analysis:algo}
\paragraph{RC4: Absence of Algorithm-Level Diversity}
cuDNN selects Winograd for $3\times3$ filters when arithmetic reduction dominates (Winograd $F(2,3)$ cuts multiply-accumulates ${\approx}2.25\times$), but neither Triton nor TileLang exposes the transformation as a schedulable primitive.
Isolating this effect on the A100 bounds its size: a deterministic A/B comparison of cuDNN on $3\times3$ stride-1 convolution differs by only 0.03\% (a ratio of 0.9997), so absent Winograd selection accounts for at most ${\approx}2$--$3\%$ of the gap.
Moreover, the gap does not track Winograd eligibility: the eligible $3\times3$ stride-1 filter (28.5\%) is no better than the ineligible stride-2 case (32.1\%), and the deepest gaps fall on $5\times5$ (17.8\%) and $7\times7$ (13.1\%), so the residual is general cuDNN implicit-GEMM tuning, not algorithm selection.

\label{sec:analysis:summary}

\begin{table*}[t]
  \centering
  \caption{Root-cause summary. Contribution = latency speedup recovered by the corresponding mitigation (\cref{sec:mitigation}); ``---'' = no direct mitigation experiment. RC0a/RC0b/RC3 are TileLang-specific; RC1/RC2/RC4 affect both DSLs; counter values agree across architectures.}
  \label{tab:rootcauses}
  \scriptsize
  \begin{tabular}{p{4.1cm}p{2.9cm}p{1.9cm}p{4.0cm}}
    \toprule
    Root cause                     & Affected kernels              & Contribution & Diagnostic counter              \\
    \midrule
    RC0a: \texttt{T.serial} instead of \texttt{T.reduce} & All TileLang reductions, norm & $1347\times$ (LayerNorm, A100) & \texttt{warp\_stall\_long\_scoreboard} (barrier ${\approx}0$) \\
    RC0b: Absent vectorized loads / dtype cast            & TileLang norm                 & ---                     & \texttt{l1tex} bytes/sector     \\
    RC1: Strided access, scalar LD  & Conv2d ($K > 1$), Triton      & $2.2\times$ with RC2    & Load bytes/sector eff.\ (12.55\%) \\
    RC2a: Auto-tuning mismatch      & Matmul, Conv2d, Depthwise     & $1.37\times$ (Matmul)   & TFLOPS vs.\ swept configs       \\
    RC2b: L2 cache residency        & Matmul $\geq 16384^2$         & ---                     & L2 hit rate, DRAM throughput    \\
    RC3: TileLang LayerNorm spill   & TileLang LayerNorm (large)    & ---                     & \texttt{local\_op} spill / occupancy (A100: 51.5\,GB ld, 254 regs, 12.4\% occ) \\
    RC4: No Winograd                & Conv2d $3\times3$             & ${\approx}2$--$3\%$     & SM utilization delta            \\
    \bottomrule
  \end{tabular}
\end{table*}

\medskip\noindent\textbf{Answer to RQ2: The hidden gap has no single cause but a small, fix-locus-separated taxonomy of them. The most dramatic gaps---normalization and reduction collapses of up to $1347\times$---are user-space authoring artifacts (RC0a). A second tier is compiler code generation: absent vectorization (RC0b, RC1), default-grid tuning coverage (RC2a), and register spill (RC3). The remainder is library maturity (RC2b, RC4), which bounds what any current DSL kernel reaches. Each cause has a distinct counter signature (\cref{tab:rootcauses}), making the taxonomy a diagnostic checklist.}

%% file: tex/mitigation.tex
This section answers RQ3: \emph{absent a comprehensive benchmark, can lightweight evaluation heuristics and a small set of recurring optimization patterns reliably flag performance-poor DSL kernels and guide their repair?}
RQ1 showed that correctness gates admit slow kernels and that a comprehensive benchmark is infeasible; RQ2 explained why the gaps arise.
We distill that evidence into a two-part \emph{evaluation heuristic} for deciding whether a kernel is efficient (\cref{sec:mitig:heuristics}) and the \emph{optimization patterns} that repair the gaps it flags (\cref{sec:mitig:norm} onward).
The optimization campaigns (\cref{sec:meth:optimization}) ran on a separate development GPU; their kernels are re-timed here on the A100 and GH200.
The central result is a clean dichotomy: most dramatic gaps are \emph{authoring artifacts} that one dominant pattern fully repairs, a minority are \emph{genuine residuals} that survive best-effort optimization, and the heuristic tells the two apart without a ground-truth benchmark.

\subsection{Evaluation Heuristics: Is This Kernel Efficient?}
\label{sec:mitig:heuristics}

We propose two complementary checks a developer can apply without a curated benchmark.

\paragraph{A comparability screen (pragmatic, baseline-relative)}
An efficient DSL kernel should (i) come within a small factor of the vendor/PyTorch baseline on representative shapes, (ii) be at least at parity---ideally faster---at large input sizes where launch and framework overheads amortize, and (iii) show no catastrophic per-shape collapse.
This screen is cheap and catches gross failures: the idiomatic TileLang LayerNorm that runs more than $300\times$ slower (\cref{sec:eval:passesslow}) fails it immediately.
Its limitation is circularity: PyTorch is both the baseline and the de-facto definition of ``good,'' so the screen cannot certify a kernel that merely matches an already-slow baseline, and it can credit a kernel that beats an \emph{unfused} eager path for the wrong reason (the RMSNorm $873\%$ ``win'' in \cref{tab:summary}, a fusion artifact); it is a screen, not a certificate.

\paragraph{A roofline anchor (baseline-independent)}
To break that circularity we anchor quality to the hardware rather than to PyTorch: for a kernel whose essential work is $W$ (bytes moved if memory-bound, FLOPs if compute-bound) and whose optimized time is $t$, the achieved roofline fraction is $\rho = W / (t \cdot P)$, where $P$ is the relevant hardware peak~\cite{williams2009roofline}.
TritonBench reports an analogous achieved-peak metric~\cite{li2025tritonbench}; we use it as a \emph{decision signal}, not a leaderboard score: a kernel near the roof is efficient \emph{regardless} of what the library does.
\Cref{tab:roofline} reports $\rho$ for the optimized kernels on the A100 (clock-locked) and the GH200 (unlocked), alongside their PyTorch-relative efficiency $E_\text{lib}$ and the \emph{cliff}---the naive-to-optimized speedup that measures how much authoring headroom each kernel hid.
The anchor does two things the comparability screen cannot.
First, it confirms the genuine residuals are genuinely far from the hardware: optimized Triton Conv2d reaches only $\rho=0.16$ and index-returning argmax only $\rho=0.10$ on the GH200, so their shortfall is real headroom, not merely a fast baseline.
Second, it de-inflates baseline-relative outliers: the RMSNorm ``$13\times$ win'' over PyTorch sits at $\rho=0.69$---efficient, but hardly beyond what the hardware allows.
On the GH200, the recovered normalization and reduction family lands at $\rho=0.41$--$0.87$, while the residual convolution and argmax kernels sit at $\rho\le0.28$ even after best-effort optimization; the A100 mirrors the split ($\rho=0.42$--$0.89$ recovered, $\le0.34$ residual), so the classification is architecture-independent.
The cliff column shows the two signals measure different axes: LayerNorm hid a $1541\times$ authoring cliff on the GH200 ($380\times$ on the A100) that the dominant pattern fully recovers, whereas Conv2d's $8.2\times$ cliff still leaves it at $\rho=0.28$, a structural ceiling.
We present the two checks together because neither suffices alone, and they disagree exactly in a judgment band near $\rho\approx0.5$: Softmax and Matmul both achieve $\rho=0.47$ on the GH200, yet Softmax is at parity ($E_\text{lib}=0.87$) while Matmul trails cuBLAS ($E_\text{lib}=0.68$)---only the two checks together make the call.

\begin{table*}[t]
  \centering
  \caption{Evaluation heuristics on the A100 (sm\_80, clock-locked) and GH200 (sm\_90, unlocked): optimized-kernel roofline fraction $\rho$ (achieved/hardware peak; \emph{mem}=HBM bandwidth, \emph{comp}=FP16 Tensor-Core), PyTorch-relative efficiency $E_\text{lib}$, and the \emph{cliff} (naive/optimized speedup). High recovered $\rho$ = authoring artifact; low $\rho$ surviving optimization = structural residual; the two architectures agree on every classification.}
  \label{tab:roofline}
  \setlength{\tabcolsep}{4pt}
  \resizebox{\textwidth}{!}{%
  \begin{tabular}{@{}llrrrrrr@{\hspace{2.5em}}llrrrrrr@{}}
    \toprule
    & & \multicolumn{3}{c}{A100-SXM4 (sm\_80)} & \multicolumn{3}{c}{GH200 (sm\_90)}
    & & & \multicolumn{3}{c}{A100-SXM4 (sm\_80)} & \multicolumn{3}{c}{GH200 (sm\_90)} \\
    \cmidrule(lr){3-5}\cmidrule(lr){6-8}\cmidrule(lr){11-13}\cmidrule(lr){14-16}
    Kernel (DSL) & Bound & $\rho$ & $E_\text{lib}$ & Cliff & $\rho$ & $E_\text{lib}$ & Cliff
    & Kernel (DSL) & Bound & $\rho$ & $E_\text{lib}$ & Cliff & $\rho$ & $E_\text{lib}$ & Cliff \\
    \midrule
    \multicolumn{8}{@{}l}{\textit{Recovered family (authoring artifacts)}}
    & \multicolumn{8}{l}{\textit{Structural residuals (survive best-effort optimization)}} \\
    LayerNorm (TL)     & mem & 0.67 & 1.25 & $380\times$  & 0.59 & 1.20 & $1541\times$ & Matmul (Tr) & comp & 0.56 & 0.78 & $1.1\times$ & 0.47 & 0.68 & $1.3\times$ \\
    RMSNorm (TL)       & mem & 0.70 & 10.2 & $325\times$  & 0.69 & 13.0 & $1520\times$ & Conv2d (TL) & comp & 0.34 & 0.60 & $6.4\times$ & 0.28 & 0.63 & $8.2\times$ \\
    MeanReduction (TL) & mem & 0.89 & 1.04 & $13.7\times$ & 0.87 & 0.97 & $13.9\times$ & Conv2d (Tr) & comp & 0.24 & 0.42 & $1.5\times$ & 0.16 & 0.34 & $2.5\times$ \\
    BatchedMatmul (TL) & mem & 0.82 & 0.94 & $23.9\times$ & 0.86 & 1.11 & $20.2\times$ & Argmax (TL) & mem  & 0.15 & 0.27 & $4.5\times$ & 0.10 & 0.22 & $3.8\times$ \\
    MaxReduction (Tr)  & mem & 0.69 & 1.13 & $8.8\times$  & 0.59 & 1.20 & $6.4\times$  & & & & & & & & \\
    MaxReduction (TL)  & mem & 0.42 & 0.68 & $12.8\times$ & 0.41 & 0.84 & $16.2\times$ & & & & & & & & \\
    LogSoftmax (TL)    & mem & 0.55$^\S$ & 0.71$^\S$ & ---$^\S$ & 0.58 & 0.90 & $5.2\times$ & & & & & & & & \\
    Softmax (TL)       & mem & 0.55$^\S$ & 0.86$^\S$ & ---$^\S$ & 0.47 & 0.87 & $4.0\times^\ddagger$ & & & & & & & & \\
    \bottomrule
  \end{tabular}%
  }

  \vspace{0.3em}
  \parbox{\textwidth}{\scriptsize TL\,=\,TileLang, Tr\,=\,Triton.
  $^\ddagger$ The naive Softmax falls back to \texttt{torch.softmax} at this shape, so its cliff is not a pure authoring comparison; the optimized $\rho$ and $E_\text{lib}$ are unaffected.
  $^\S$ A100 Softmax/LogSoftmax report the streaming variant (\cref{sec:mitig:summary}); the in-tree full-row-cache variant collapses occupancy ($\rho{=}0.006$/$0.028$), so its cliff does not transfer and is omitted.
  \texttt{logsumexp} (omitted) register-spills on \texttt{sm\_90} but \emph{not} \texttt{sm\_80} (A100: $0$ spill, $93$~regs, $E_\text{lib}{=}582\%$)---an sm\_90-specific RC3 residual; see \cref{sec:mitig:other,tab:mitigation}.}
\end{table*}

\subsection{Normalization Kernels: Correcting RC0}
\label{sec:mitig:norm}

\medskip\noindent\textbf{Finding: Two user-space fixes---\texttt{T.serial}$\to$\texttt{T.reduce} (dominant, RC0a) and native bf16/fp16 I/O without intermediate \texttt{.float()} casts---bring TileLang LayerNorm to $124\%$ and RMSNorm to $990\%$ of PyTorch on the A100 ($1347\times$/$1157\times$ over the unoptimized baseline), confirming RC0 as the primary, user-correctable cause of the normalization deficit.}

The dominant fix replaces the \texttt{T.serial} loop with a single \texttt{T.reduce} (subsequent iterations confirmed the loop structure, not thread configuration, was the bottleneck); native bfloat16 I/O, removing intermediate \texttt{.float()} casts, then brings LayerNorm from 364~ms to 0.270~ms and RMSNorm to 0.254~ms (the latter exceeds 100\% because the fixed kernel fuses the scaling pass PyTorch's eager two-pass path does not)---near the $\approx$0.18~ms bandwidth floor for an $8192^2$ bfloat16 tensor.
This is not a new technique; the contribution is showing RC0 fully explains the anomaly and is user-correctable without algorithmic change (the RC0b codegen deficiency is sidestepped, not fixed).
The 18-iteration LayerNorm search trajectory is in \appref{app:repro:trajectory}.

\subsection{Convolution: Partial Recovery via Implicit GEMM}
\label{sec:mitig:conv}

\medskip\noindent\textbf{Finding: Restructuring Conv2d as an FP16 Tensor-Core implicit GEMM with aligned padding (RC1) and an expanded autotune space (RC2) reaches 36--77\% of cuDNN across $1\times1$--$7\times7$ on the A100---narrowing but not closing the gap.}

The restructuring unfolds the convolution into an on-the-fly implicit GEMM~\cite{zhou2021implicit} that FP16 Tensor Cores accelerate, with 16-byte input padding enabling \texttt{LDG.128} loads (RC1) and an expanded autotune space (smaller \texttt{BLOCK\_K}, more pipeline stages) covering configurations absent from the default list (RC2).
Of the residual $\approx$20\% gap, absent Winograd selection contributes only $\approx$2--3\% (a 0.03\% deterministic-vs-non-deterministic delta for $3\times3$ stride-1); the remainder reflects general cuDNN implicit-GEMM tuning---kernel selection, split-K, shared-memory tile sizing---that the single-lowering DSL kernel does not match (RC4).
Per-filter optimized efficiencies range from 77.4\% at $1\times1$ to 36.1\% at $7\times7$ (all correct); closing this tuning gap and adding in-DSL Winograd support are future work.
The depthwise variant ($\le$5.5\% in \cref{sec:evaluation}) is outside this rewrite's scope: its grouped lowering degenerates into 512 per-group GEMM launches whose cost is launch- and code-generation-bound (RC1), and none of our patterns repair it. We flag it as the clearest open convolution gap.

\subsection{GEMM and Reduction Kernels}
\label{sec:mitig:other}

Beyond the normalization kernels, we optimized the GEMM kernel and the full TileLang
reduction/softmax family to test how broadly---and how completely---the RQ1 gaps recover.

\textbf{Square matmul (Triton, FP16).}
Adding \texttt{@triton.autotune} with an expanded set of tile configurations
and a \texttt{GROUP\_SIZE\_M} L2 cache swizzle (which reorders output tiles to improve L2 data reuse across the $M$-dimension)
reduces latency from 1.08~ms to 0.79~ms ($1.37\times$, $E_\text{lib} = 77\%$)
on a $4096\times4096$ matmul, re-timed on the A100;
the same expanded search recovers $E_\text{lib}$ from $59.7\%$ to $81.9\%$ ($1.37\times$) at the RQ1 $16384^2$ shape.
Further iterations (persistent kernel, \texttt{max\_num\_imprecise\_acc}) converged without gain, so the expanded configuration set and L2 swizzle are the effective ceiling---confirming RC2: the gain is a property of the \emph{expanded} search space (\cref{sec:eval:tuning}), not of a different problem shape.
(\Cref{tab:roofline} also lists optimized TileLang Conv2d and BatchedMatmul kernels from the same protocol (\cref{sec:meth:optimization}); their campaigns are omitted for space.)

\textbf{Reduction and softmax family (TileLang).}
The same RC0 fix---replacing \texttt{T.serial} reduction loops with native \texttt{T.reduce}
and streaming the reduction dimension in tiles (an online, max-rescaled pass for the softmax
variants) rather than materializing the full row in a fragment---generalizes across the
\emph{entire} TileLang reduction and normalization family on the A100 (\cref{tab:mitigation}). Every catastrophically slow kernel recovers to
PyTorch-comparable or better: \texttt{max\_reduction}, \texttt{mean\_reduction}, and
\texttt{logsumexp} \emph{exceed} PyTorch ($132\%$, $99\%$, $582\%$), and \texttt{softmax} and
\texttt{log\_softmax} reach $86\%$ and $71\%$---all numerically correct, with $4$--$29\times$
speedups over the idiomatic kernels. The lone within-A100 exception is
index-returning \texttt{argmax}: tiled shared-memory bulk loads and native FP16 I/O (addressing RC0 and RC1) improve it $5.2\times$ (11.97~ms to 2.31~ms), yet it stops at $27\%$ because \texttt{torch.argmax} (0.62~ms) is already well tuned---and the \emph{value-only} reduction over
the identical shape (\texttt{max\_reduction}) reaches $132\%$, isolating the residual to the index pass.
Re-timing the same optimized kernels on the GH200 confirms the pattern transfers
(LayerNorm $120\%$, the rest of the family $84$--$1303\%$; \cref{tab:mitigation,tab:roofline}),
with one instructive exception: \texttt{logsumexp}, whose
A100-tuned wide fp32 fragment register-spills on \texttt{sm\_90} (255~registers, ${\sim}280$~GB of
local-memory spill traffic, $12\%$ occupancy by Nsight Compute), collapsing to $1.0\%$. The spill is an RC3-class effect, not
the RC0 reduction idiom: retuning the fragment width recovers it only to $63\%$, so on the GH200
\texttt{logsumexp} is a genuine residual---an ``optimized'' kernel validated on one GPU that is
$100\times$ off on another with no correctness signal, precisely the evaluation gap a single-target benchmark cannot close.

\subsection{Summary and Remaining Gap}
\label{sec:mitig:summary}

\Cref{tab:mitigation} summarizes the per-category results (A100, with GH200 efficiency); they separate into two kinds---the central result of RQ3.
The TileLang \emph{normalization and reduction family} are \emph{authoring artifacts}: the RC0 fix recovers every family kernel to PyTorch-comparable or better, closing gaps as large as $300\times$, with GH200 \texttt{logsumexp} the lone exception---so even a fixed kernel carries a per-target tuning that must be re-validated.
The \emph{genuine residuals}---Conv2d ($42\%$), large square GEMM ($77\%$), and index-returning Argmax ($27\%$)---survive best-effort optimization because cuBLAS, cuDNN, and \texttt{torch.argmax} are themselves well-tuned on the data-center A100; of the Conv2d gap only ${\approx}2$--$3\%$ is absent Winograd selection (RC4).

\paragraph{Cross-architecture portability is itself an authoring concern}
A \emph{correct, hand-optimized} DSL kernel can pass on one GPU yet collapse on another, so single-GPU or single-shape benchmarking cannot certify it.
\texttt{logsumexp} is one direction (fast on the A100, register-spilling on the GH200); the in-tree TileLang Softmax is the other, purely an authoring choice: it caches the entire row in shared memory, so its per-block footprint erodes A100 occupancy---$20\times$ slower than PyTorch at $N{=}8192$ and $111\times$ at $N{=}32768$, while numerically correct---yet the identical kernel hits parity on the GH200, whose larger shared-memory budget hides the defect.
The streaming variant (tiling the reduction over fixed $4$~KB blocks) holds parity across the sweep ($\rho{=}0.55$, \cref{tab:roofline}), so ``stream the reduction; do not cache the full row'' is a portable authoring pattern.

\begin{table}[t]
  \centering
  \caption{Summary of mitigation results. A100-SXM4 latencies in ms (lower is better);
    $E_\text{lib}$ = optimized-kernel efficiency relative to PyTorch/cuDNN on each GPU.
    \textbf{Bold} = ${\geq}95\%$ library efficiency on that GPU.}
  \label{tab:mitigation}
  \resizebox{\columnwidth}{!}{
  \begin{tabular}{lrrrrrl}
    \toprule
    & \multicolumn{3}{c}{A100-SXM4 (ms)} & \multicolumn{2}{c}{$E_\text{lib}$} & \\
    \cmidrule(lr){2-4}\cmidrule(lr){5-6}
    Kernel & PyTorch & Before & After & A100 & GH200 & Root cause \\
    \midrule
    \multicolumn{7}{l}{\textit{Triton}} \\
    \quad Matmul    & 0.607 & 1.081 & 0.788          & 77.0\%            & 68\%   & RC2 \\
    \quad Conv2d    & 3.44 & 17.86 & 8.12           & 42.4\%            & 34\%   & RC1+RC2 \\
    \midrule
    \multicolumn{7}{l}{\textit{TileLang}} \\
    \quad LayerNorm & 0.335 & 364  & \textbf{0.270} & \textbf{124\%}    & \textbf{120\%} & RC0 \\
    \quad RMSNorm       & 2.52  & 294   & \textbf{0.254} & 990\%$^\dagger$  & 1303\%$^\dagger$ & RC0 \\
    \quad Softmax       & 0.539 & 2.86  & 0.626          & 86.1\%           & 87\%   & RC0 \\
    \quad LogSoftmax    & 0.442 & 2.65  & 0.624          & 70.8\%           & 90\%   & RC0 \\
    \quad MaxReduction  & 0.560 & 11.97 & \textbf{0.418} & \textbf{131.6\%} & 84\%   & RC0 \\
    \quad MeanReduction & 0.768 & 10.65 & \textbf{0.766} & \textbf{99.5\%}  & \textbf{97\%} & RC0 \\
    \quad LogSumExp     & 2.40  & 4.18  & \textbf{0.412} & \textbf{581.7\%} & 1.0\%$^\ddagger$ & RC0/RC3 \\
    \quad Argmax        & 0.622 & 11.97 & 2.306          & 27.0\%           & 22\%   & RC0+RC1 \\
    \bottomrule
  \end{tabular}
  }
  \vspace{0.3em}
  {\footnotesize $^\dagger$ RMSNorm $E_\text{lib}>100\%$: the fixed kernel fuses the scaling pass PyTorch's eager path leaves unfused.
  $^\ddagger$ LogSumExp's A100-tuned config register-spills on sm\_90 but not sm\_80; the $1.0\%$/$581.7\%$ split, retuned $63\%$ ceiling, and RC3 attribution are in \cref{sec:mitig:other,tab:roofline}.
  \emph{Before} = unoptimized DSL kernel (Matmul: plain Triton at $4096^2$; Conv2d: baseline vs.\ implicit-GEMM Triton at $3\times3$); norm/reduction \emph{Before} kernels are memory-latency-bound and clock-invariant. GH200 $E_\text{lib}$ from \cref{tab:roofline} (unlocked).}

\end{table}
All \cref{tab:mitigation} mitigations are re-timed on the A100 under locked clocks (matmul: the $4096^2$ autotune value, reconciled with $16384^2$ in \cref{sec:mitig:other}) and revalidate against the same per-dtype tolerances (5/5 pass, 0 edge-case crashes).

%% file: tex/discussion.tex
\subsection{Implications for DSL Developers}
\label{sec:disc:dsl}

Our results point to three compiler-side directions, each tagged by the root cause it addresses.
\textbf{Vectorize convolution accesses (RC1):} extend Triton's alias/layout analysis to emit wide \texttt{LDG.128} loads for strided spatial patterns---a code-generation fix, not an API change.
\textbf{Make auto-tuning shape-aware (RC2):} RC2a is user-space-fixable today by expanding the configuration list (\cref{sec:mitigation}), but the residual RC2b calls for search adaptive to operator shape, as sketch-based methods like Ansor~\cite{zheng2020ansor} demonstrate for irregular operators. Tuning at the deployment shape is itself costly ($211$\,s vs.\ sub-second, \cref{sec:eval:survey}), reinforcing that exhaustive tuned coverage is infeasible and the heuristics of \cref{sec:mitigation} are the practical screen.
\textbf{Support algorithmic alternatives (RC4):} exposing Winograd as a schedulable primitive would let the compiler choose execution strategy by filter shape, though the benefit is small ($\approx$2--3\%).

\subsection{Implications for Practitioners and Benchmark Designers}
\label{sec:disc:practitioners}

\textbf{Performance depends strongly on operator class.}
Triton is near parity on element-wise and normalization workloads, a favorable programmability trade-off; GEMM carries a cost that may be acceptable when fusion or customization is required; convolution is the weakest category for both DSLs and must be validated against a library baseline, not treated as a drop-in replacement.

\textbf{The root-cause taxonomy guides debugging.}
For TileLang reductions and normalization, replace \texttt{T.serial} with \texttt{T.reduce} (RC0a) and inspect register spill (RC3); for convolution, check \texttt{LDG.128} vectorization (RC1) and whether the tuning space is under-populated (RC2a) or needs a broader strategy (RC2b); Winograd (RC4) contributes only $\approx$2--3\%.
This checklist mirrors \cref{sec:analysis,sec:mitigation}.

\textbf{Benchmarks should gate on performance, not just correctness.}
The passes-but-slow result (\cref{sec:eval:passesslow}) yields a concrete recommendation: a benchmark that reports speedup without gating on performance certifies the wrong property. A baseline-independent roofline check (\cref{sec:mitig:heuristics}) is a practical acceptance criterion that needs no curated fast reference and would have flagged every passes-but-slow kernel in our study.

\subsection{Limitations and Threats to Validity}
\label{sec:disc:limitations}

\textbf{Kernel authorship (construct validity).}
All TileLang suite kernels are our re-implementations (\cref{sec:meth:kernels}), so the representativeness of the naive kernels---including the headline $300\times$ LayerNorm---is a threat: a different author might not write the \texttt{T.serial} idiom.
Two observations bound it: the same idiom, written independently across the whole reduction and normalization family, produces the same class of gap; and the in-tree TileLang softmax---authored upstream, not by us---exhibits the same passes-but-slow failure mode (full-row caching, up to $111\times$ slower while numerically correct; \cref{sec:mitig:summary}), so the phenomenon is not an artifact of our authorship.

\textbf{Baselines and measurement (construct and internal validity).}
Library efficiency is measured against the strongest library path we could exercise (library-backed PyTorch calls, \texttt{cudnnFind} selection, a large cuBLAS workspace, algorithms recorded); a stronger unexercised baseline would only widen the gaps.
Locked-clock re-timing over 100 runs shows 0.0--0.9\% run-to-run relative standard deviation and NCU counters are stable within 2\% across five runs, so the gaps lie far outside measurement variation. Auto-tuning at small shapes can pick configurations arbitrary at deployment shapes; the three affected kernels were re-tuned at the deployment shape (\appref{sec:threats}).

\textbf{Operator and hardware scope (external validity).}
We evaluate forward-pass kernels only; backward kernels may expose bottlenecks not captured here.
Our primary measurements span the A100 (Ampere, sm\_80) and GH200 (Grace Hopper, sm\_90), with cross-architecture replay on the A100-PCIE-40GB and H100-80GB-HBM3 confirming the gaps generalize across form factors and GPU families (\appref{sec:xarch}).
The results characterize NVIDIA datacenter platforms; ROCm and Intel XPU backends are future work.

\textbf{Heuristic assumptions and conclusion validity.}
The roofline anchor depends on an essential-work model and an assumed hardware peak; mis-estimating either shifts $\rho$ (and the two checks disagree near $\rho\approx0.5$, \cref{sec:mitig:heuristics}), so we use $\rho$ as a decision signal, not a precise figure.
Our root-cause claims combine measurements, counters, and targeted mitigations, which strengthen the causal reading but do not rule out unisolated factors.
The 22-kernel suite is a diagnostic probe, not a proposed benchmark---a central claim of this work is that a comprehensive benchmark is infeasible.
Finally, both DSLs are evolving; measurements reflect the pinned versions in \cref{sec:meth:setup}, and our released suite supports re-evaluation.

%% file: tex/related_work.tex
\subsection{GPU Kernel DSLs and Performance Tooling}
\label{sec:rel:dsls}

Halide~\cite{ragan-kelley2013halide} introduced the separation of algorithm
from schedule, a principle that informs TVM~\cite{chen2018tvm} and
TileLang~\cite{wang2025tilelang}; Triton~\cite{tillet2019triton} instead infers
shared-memory layout, synchronization, and pipelining from an implicit tile
abstraction, and its \texttt{torch.compile} integration~\cite{li2023torchcompile}
has made it the most widely deployed GPU DSL.
ThunderKittens~\cite{spector2024thunderkittens} instead exposes warp-level tile
operations as a thin C++ header library; we characterize performance at the DSL level.
TVM's auto-scheduler Ansor~\cite{zheng2020ansor} shows that hierarchical program
search outperforms template-guided auto-tuning for irregular shapes such as
convolution---consistent with RC2 in \cref{sec:analysis}---and MLIR-based
pipelines~\cite{bik2022mlir,vasilache2022composable} generate near-peak GEMM and
fused-attention code within compiler infrastructure.

TritonForge~\cite{li2025tritonforge} proposes a profiling-guided LLM loop for
automated Triton kernel optimization; its finding that coalescing failures and
low occupancy dominate kernel-level inefficiency is consistent with RC1 and
RC3; it automates fix generation, whereas we characterize systematically across a taxonomy.
The vendor libraries themselves---cuDNN~\cite{chetlur2014cudnn},
cuBLAS~\cite{cublas2024}, and CUTLASS~\cite{cutlass2023}---are extensively
engineered, but systematic DSL-versus-library comparisons at this scale and
granularity are absent from the literature.

\subsection{Benchmarking GPU Software}
\label{sec:rel:bench}

The benchmarks closest to our setting evaluate DSL and LLM-generated kernels.
TritonBench~\cite{li2025tritonbench} evaluates Triton kernel generation by LLMs,
measuring functional correctness and GPU efficiency as a fraction of hardware peak---
a roofline-anchored metric that predates and motivates the anchor we adopt in \cref{sec:mitig:heuristics}, and which we build on rather than introduce.
KernelBench~\cite{kernelbench} is the de-facto benchmark for LLM kernel generation,
but it gates on correctness alone and reports speedup only as an outcome.
Its single reward-hacking guard flags only implausibly \emph{fast} kernels (motivated by real incidents of generators exploiting evaluation harnesses to fake speedups~\cite{sakana2025aicuda}), so a merely slow kernel is never rejected.
Both benchmarks evaluate only a narrow slice of the (DSL, data type, shape, hardware) space, and MLPerf Inference~\cite{reddi2020mlperf} benchmarks end-to-end throughput while treating the kernel stack as a black box.
We do not propose another benchmark: we show the prevailing ones admit performance-poor kernels (\cref{sec:evaluation}), characterize the hidden gap with hardware counters, and distill heuristics and optimization patterns that guide development without one.

\subsection{Efficiency-Aware Evaluation of Generated Code}
\label{sec:rel:se}

A parallel line of software-engineering work makes the correct-but-slow argument for CPU code.
EvalPerf~\cite{liu2024evalperf}, Mercury~\cite{du2024mercury}, EffiBench~\cite{huang2024effibench},
and ENAMEL~\cite{qiu2024enamel} show that LLM-generated solutions that pass functional tests can be
far less efficient than expert code, and each scores efficiency directly (differential performance,
runtime-percentile, and \texttt{eff@k} metrics).
These benchmarks target sequential CPU programs, where a canonical reference and input generator
suffice; GPU kernels have no canonical reference beyond the vendor library they may be designed to
out-specialize, efficiency depends on a combinatorially infeasible (data type $\times$ shape $\times$
architecture) space (\cref{sec:evaluation}), and we attribute each gap to a fix locus rather than scoring it.
Classic performance-bug studies~\cite{jin2012understanding,nistor2013discovering} document defects
that pass functional tests while degrading performance, and the test-oracle
literature~\cite{barr2015oracle} frames our observation precisely: a correctness oracle alone is an
inadequate oracle for replacement quality.
Closest to our setting, Rathnasuriya \etal~\cite{rathnasuriya2026tilebugs} study 301 tile-program code-generation bugs and find that they surface as silent correctness \emph{or performance} faults rather than crashes---evidence, from the defect side, for the oracle gap we quantify from the performance side.

%% file: tex/conclusion.tex
We studied how to \emph{evaluate} DSL GPU kernels: can a developer or an automated generator tell whether a correct Triton or TileLang kernel is performant? The benchmarks practitioners rely on gate only on correctness---a naive but idiomatic TileLang kernel passes KernelBench while running more than $300\times$ slower than PyTorch---and a comprehensive benchmark is combinatorially infeasible, so we characterized the hidden gap across 22 kernels in five operator categories and distilled it into practical guidance.

Most of the gap is an \emph{authoring} artifact: correcting a single TileLang idiom (\texttt{T.serial}$\to$\texttt{T.reduce} with native-dtype I/O) recovers the normalization and reduction family to library-comparable performance or better on both GPUs, with two diagnosed exceptions---index-returning \texttt{argmax} (27\%) and, on the GH200 only, \texttt{logsumexp} (register spill). The remainder---convolution and large GEMM---is a \emph{genuine residual} where even a best-effort DSL kernel trails cuDNN/cuBLAS. We separate causes by fix locus and validate two lightweight heuristics---a comparability screen and a roofline anchor---that together tell efficient kernels from poor ones.

A correct DSL kernel is not yet a fast one, and existing benchmarks do not close that gap; until they do, the heuristics and optimization patterns here give developers---and the kernel-generating tools increasingly producing DSL code---a practical way to judge and improve kernel quality.

%% file: tex/appendix.tex

\section{Threats to Validity}
\label{sec:threats}

\textbf{Construct validity.}
Our primary metric, library efficiency, measures DSL performance relative to the strongest library baseline we were able to exercise for each kernel. Because cuBLAS and cuDNN select among multiple internal algorithms, any failure to invoke their best-performing configuration would make the DSL appear closer to library performance than it actually is. To reduce this risk, we use library-backed PyTorch paths consistent with standard practice, invoke \texttt{cudnnFind} for convolution, record the selected algorithms, and allow cuBLAS to use a large workspace.
We report under default cuDNN flags (no explicit \texttt{cudnn.benchmark} toggle, no NHWC conversion); convolutions run in the PyTorch-default NCHW layout, and algorithm-selection variance is bounded by \texttt{cudnnFind} defaults.

\textbf{Internal validity.}
Profiling-based analysis may be affected by measurement noise and tool overhead. We mitigate this threat by separating end-to-end timing from counter collection, using repeated runs for both, and verifying that key Nsight Compute measurements remain stable across repetitions. All experiments are executed in isolation on a fixed software and hardware setup to reduce confounding effects from concurrent workloads or environmental variation. Locked-clock re-timing (graphics 1215 MHz / memory 1215 MHz) over 100 runs shows run-to-run relative std-dev of 0.0--0.9\% on near-parity kernels, so the reported gaps lie far outside measurement variation rather than being artifacts of frequency variation.

\textbf{Auto-tuning shape sensitivity.}
Our auto-tuning sweep selects each kernel's configuration by timing the candidate grid at a fixed per-kernel shape that, for several kernels, is smaller than the large deployment shape used for benchmarking.
At the smaller shape the candidates often fall within measurement noise, so the selected configuration can be effectively arbitrary and is occasionally slower than the implementation's hardcoded default at the deployment shape (we observed Triton \texttt{mean\_reduction} at $+136\%$ and \texttt{batched\_matmul} at $+54\%$ relative to the default).
We re-tuned the kernels where this occurred (\texttt{batched\_matmul}, \texttt{mean\_reduction}, and \texttt{attention} in Triton) at the deployment shape, restoring configurations at or below the default ($1.3$--$3.0\times$ faster than the small-shape selection); this touches only those auto-tuned rows and does not affect the reported root causes or heuristics.

\textbf{External validity.}
Our study covers 22 kernels across five operator categories, but it does not represent the full design space of GPU workloads. In particular, the evaluated configurations emphasize common deep learning operators and shapes rather than uncommon cases such as highly dilated convolutions, extreme group counts, or very small batches. Conv2d is evaluated across 1$\times$1, 3$\times$3, 5$\times$5, 7$\times$7, depthwise, and strided variants.
In addition, our primary measurements span two architectures---the A100-SXM4-40GB (Ampere, sm\_80) and the GH200 (Grace Hopper, sm\_90)---with further cross-architecture replay on the A100-PCIE-40GB (same sm\_80, a form-factor robustness check) and H100-80GB-HBM3 (a cross-family generalization check); results on other vendor backends may still differ.

\textbf{Conclusion validity.}
Our root-cause claims are based on a combination of performance measurements, hardware-counter evidence, and targeted mitigations. Although the mitigation results strengthen the causal interpretation, they do not rule out additional compiler or runtime factors that were not isolated in this study. The conclusions should therefore be read as evidence-supported explanations for the observed gaps, rather than as an exhaustive account of all possible causes.

\textbf{The suite is a probe, not a proposed benchmark.}
Our 22-kernel suite is deliberately not offered as a comprehensive performance benchmark---a central claim of this work (\cref{sec:evaluation}) is that such a benchmark is combinatorially infeasible to build and maintain. The suite instead functions as a diagnostic \emph{probe}: it is large enough to expose the evaluation gap (correct kernels that existing benchmarks admit yet run far below the library), to root-cause that gap across operator classes, and to validate the proposed heuristics, but it is not intended to certify any kernel as production-ready. The contribution is the evaluation methodology---the passes-but-slow demonstration, the authoring/code-generation/library-maturity taxonomy, and the comparability-plus-roofline heuristics---rather than the suite's size.

\textbf{Heuristic assumptions.}
The roofline anchor depends on an essential-work model (bytes moved or FLOPs) and an assumed hardware peak for each kernel; mis-estimating either shifts the achieved fraction $\rho$, and the two checks disagree in a judgment band near $\rho\approx0.5$ (\cref{sec:mitig:heuristics}). We therefore use $\rho$ as a decision signal alongside the comparability screen rather than as a precise efficiency figure, and we record the peak assumptions with each measurement (\cref{tab:roofline}). The suite magnitude is measured on the A100 and the evaluation-gap and roofline results on the GH200---two primary architectures carrying complementary evidence---with per-architecture magnitudes reported where they differ.


\section{Cross-Architecture Generalization}
\label{sec:xarch}

The main-paper tables report results on the two primary architectures (A100-SXM4 and GH200).
As a robustness check, we replay the kernel suite and the root-cause taxonomy along an A100-SXM4 / A100-PCIE / H100 axis: the A100-SXM4 serves as the anchor, the A100-PCIE-40GB is a same-architecture (\texttt{sm\_80}) form-factor control, and the H100-80GB-HBM3 is a cross-family (\texttt{sm\_90}) control. The GH200 primary results remain in the main-paper tables.

\begin{figure}[t]
  \centering
  \IfFileExists{figures/overview_efficiency_gh200.pdf}{%
    \includegraphics[width=\linewidth]{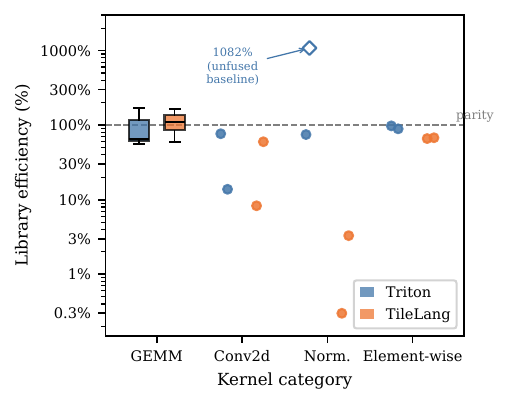}%
  }{%
    \fbox{\parbox[c][6cm][c]{\linewidth}{\centering
      \textit{[Figure: GH200 library efficiency (\%) by kernel category and DSL.
      Generate with \texttt{python3 figures/gen\_overview\_gh200.py}.]}}}%
  }
  \caption{Library efficiency (\%) of Triton and TileLang kernels relative to cuBLAS/cuDNN
    on the NVIDIA GH200-480GB, by kernel category (log scale; dashed line = 100\% parity).
    Same layout as \Cref{fig:overview}: GEMM shows IQR boxes; other categories show per-kernel dots.
    $\diamond$~1082\% (Triton, Norm.) marks \texttt{rms\_norm}: unfair-baseline artifact, not a DSL win.
    TileLang Conv2d is competitive on both (GH200 59.9\% large; A100 $\sim$59\%);
    TileLang Norm.\ remains collapsed (0.3\%).
    Attention excluded (\Cref{sec:eval:gemm}).}
  \label{fig:overview:gh200}
\end{figure}

\Cref{fig:overview:gh200} shows the per-category library-efficiency distribution on the GH200,
parallel to \Cref{fig:overview} in the main paper.
The GH200 profile is consistent with the A100: TileLang convolution is competitive on both
(GH200 small-shape 8.3\% and large-shape 59.9\%; A100 7.2\% and 59.0\%),
while TileLang normalization remains collapsed on both (LayerNorm 0.3\%). This confirms that
the normalization gap is an authoring artifact independent of architecture, while
the residual convolution gap is the Triton arm, which holds on both architectures (\Cref{sec:analysis}).

\begin{table}[t]
  \centering
  \caption{\textbf{Cross-architecture generalization.} Median library efficiency $E_\text{lib}$ (\%) per
    kernel category for each DSL on the primary A100-SXM4 (sm\_80) and the cross-architecture replays
    A100-PCIE (sm\_80, form factor) and H100 (sm\_90, Hopper). All cells are \emph{untuned defaults},
    frozen from the initial pre-tuning sweep as a like-for-like robustness check; they therefore differ
    from the tuned main-paper profile (\cref{tab:summary}). In particular, the TileLang convolution
    medians (7.8--11.1\%) reflect the untuned configuration diagnosed as a tuning artifact in
    \cref{sec:evaluation} (tuned: 59.3\% on the A100), and the Triton normalization medians include the
    \texttt{rms\_norm} unfused-baseline artifact that \cref{tab:summary} excludes. The qualitative
    profile is preserved across architectures: GEMM and element-wise competitive, Triton convolution
    and TileLang normalization severely behind.}
  \label{tab:xarch:summary}
  \resizebox{\columnwidth}{!}{
  \begin{tabular}{lccc@{\hskip 10pt}ccc}
    \toprule
    & \multicolumn{3}{c}{Triton} & \multicolumn{3}{c}{TileLang} \\
    \cmidrule(lr){2-4}\cmidrule(lr){5-7}
    Category & SXM4 & PCIE & H100 & SXM4 & PCIE & H100 \\
    \midrule
    GEMM & 77.3\% & 75.5\% & 68.4\% & 69.4\% & 60.3\% & 69.2\% \\
    Convolution & 31.6\% & 38.0\% & 37.7\% & 9.5\% & 11.1\% & 7.8\% \\
    Normalization & 121.5\% & 137.3\% & 130.5\% & 1.0\% & 0.8\% & 0.9\% \\
    Element-wise/Reduction & 65.3\% & 73.0\% & 72.9\% & 51.0\% & 55.3\% & 60.9\% \\
    \bottomrule
  \end{tabular}}
\end{table}


\begin{table}[t]
  \centering
  \caption{\textbf{Root-cause taxonomy reproduces across architectures.} Each corrected root cause
    measured on all three GPUs with the identical portable harness. All reproduce, confirming they
    are properties of the DSL/compiler rather than of a single GPU (cf.\ \cref{tab:xarch:summary}).}
  \label{tab:xarch:rootcause}
  \resizebox{\columnwidth}{!}{
  \begin{tabular}{lcccl}
    \toprule
    Root cause & SXM4 & PCIE & H100 & Reproduces \\
    \midrule
    RC0 TileLang LayerNorm anomaly ($8192^2$) $E_\text{lib}$ & 0.09\% & 0.09\% & 0.08\% & yes (memory-latency bound) \\
    Register-spill check on Triton conv ($n_\text{spills}$; RC3 does not extend to conv) & 0 & 0 & 0 & yes (no spill; occupancy-bound) \\
    RC4 Winograd upper-bound contribution & 0.1\% & 2.0\% & 20.7\%\textsuperscript{\dag} & yes ($\sim$0--2\%, not primary) \\
    FP32 \texttt{T.gemm} TF32 lowering (rel.\ err) & 633$\times$ & 633$\times$ & 1268$\times$ & yes (TF32 mantissa class) \\
    \bottomrule
  \end{tabular}}
  {\footnotesize \textsuperscript{\dag}~The cuDNN deterministic-mode (Winograd-off) timing on H100 is high-variance ($\sigma\!\approx\!27\%$ of median, un-locked clocks),
    so its determinism A/B over-states the Winograd upper bound; the stable, locked-clock A100-SXM4
    and A100-PCIE measurements bound the Winograd contribution at ${\le}2\%$.}
\end{table}

\begin{table}[t]
  \centering
  \caption{\textbf{GEMM autotuning recovery generalizes (RC2).} Triton matmul $E_\text{lib}$ before
    (plain, heuristic) and after (expanded autotune search) on all three GPUs. The
    default-grid-versus-expanded-search gap (\cref{sec:eval:tuning} vs.\ \cref{sec:mitig:other})
    is a search-space artifact on every architecture, not shape- or hardware-specific.}
  \label{tab:xarch:autotune}
  \resizebox{\columnwidth}{!}{
  \begin{tabular}{lccc@{\hskip 10pt}ccc}
    \toprule
    & \multicolumn{3}{c}{Triton plain} & \multicolumn{3}{c}{Triton autotuned} \\
    \cmidrule(lr){2-4}\cmidrule(lr){5-7}
    Shape & SXM4 & PCIE & H100 & SXM4 & PCIE & H100 \\
    \midrule
    $4096^2$ & 56.4\% & 57.7\% & 43.2\% & 76.5\% & 110.1\% & 33.1\% \\
    $16384^2$ & 32.5\% & 38.3\% & 33.6\% & 81.8\% & 87.5\% & 70.2\% \\
    \bottomrule
  \end{tabular}}
  {\footnotesize At the RQ1 $16384^2$ shape, expanded autotuning recovers Triton on all three GPUs
    (plain$\to$autotuned). The sub-millisecond $4096^2$ cells on the un-locked PCIE/H100 replays carry
    higher relative noise; the locked-clock A100-SXM4 primary is the reference measurement.}
\end{table}

\begin{table}[t]
  \centering
  \caption{\textbf{Convolution filter sweep across architectures} (large shape $32{\times}256{\times}128^2$,
    groups$=$1, stride$=$1). Triton $E_\text{lib}$ vs cuDNN for each GPU, with the measured Triton
    register-spill count. The gap widening with filter size and the absence of register spilling both
    hold across architectures.}
  \label{tab:xarch:conv}
  \begin{tabular}{lcccc}
    \toprule
    & \multicolumn{3}{c}{Triton $E_\text{lib}$} & Triton \\
    \cmidrule(lr){2-4}
    Filter & SXM4 & PCIE & H100 & $n_\text{spills}$ \\
    \midrule
    $1\times1$ & 28.7\% & 37.0\% & 24.7\% & 0 \\
    $3\times3$ & 19.3\% & 24.4\% & 11.4\% & 0 \\
    $5\times5$ & 13.7\% & 16.9\% & 13.4\% & 0 \\
    $7\times7$ & 9.8\% & 12.5\% & 15.7\% & 0 \\
    \bottomrule
  \end{tabular}
\end{table}

\section{Per-Kernel Library Efficiency}
\label{app:perkernel}

\Cref{tab:perkernel} lists $E_\text{lib} = t_\text{PyTorch} / t_\text{DSL}$ for each kernel at the large input shape on the A100-SXM4, tuned configurations.
Values below 100\% indicate the DSL is slower than the library baseline.
Four near-parity element-wise kernels (\texttt{leaky\_relu}, \texttt{swiglu}, \texttt{logsumexp}, \texttt{cross\_entropy}) exceeded 100\% on both DSLs at the large shape and are omitted from this table; their main-paper medians appear in \Cref{tab:summary}. With those four, the table accounts for all 22 suite kernels.

\begin{table}[h]
  \centering
  \caption{Per-kernel library efficiency $E_\text{lib}$ (\%) at the large input shape (A100-SXM4, tuned).
    $E_\text{lib} < 100\%$ = DSL slower than PyTorch.
    Annotated values above 100\% reflect baseline artifacts, not DSL speed advantages (see footnotes).}
  \label{tab:perkernel}
  \resizebox{\columnwidth}{!}{%
  \begin{tabular}{llrr}
    \toprule
    Kernel & Input shape & Triton & TileLang \\
    \midrule
    \multicolumn{4}{l}{\textit{GEMM}} \\
    \texttt{matmul}              & $16384\!\times\!16384$ FP16          & 59.7\%  & 78.1\% \\
    \texttt{batched\_matmul}     & $128\!\times\!2048\!\times\!2048$ FP16 & 97.9\%  & 94.2\% \\
    \texttt{linear\_activation}  & $(1,2048,4096)$ FP16                 & 185.5\%  & \textbf{468.5\%}$^{\dagger}$ \\
    \midrule
    \multicolumn{4}{l}{\textit{Attention (excluded from main comparison; see \Cref{sec:eval:gemm})}} \\
    \texttt{attention}           & QKV $(8,32,2048,128)$ FP32           & \textbf{7691\%}$^{\ddagger}$ & 54.3\% \\
    \midrule
    \multicolumn{4}{l}{\textit{Convolution}} \\
    \texttt{conv2d}              & $32\!\times\!256\!\times\!128^2$, $3\!\times\!3$ FP16 & 28.6\% & 59.0\% \\
    \midrule
    \multicolumn{4}{l}{\textit{Normalization}} \\
    \texttt{layer\_norm}         & $8192\!\times\!8192$ BF16            & 90.2\%  & 0.3\%  \\
    \texttt{rms\_norm}           & $8192\!\times\!8192$ FP16            & \textbf{873\%}$^{\S}$ & 3.2\% \\
    \midrule
    \multicolumn{4}{l}{\textit{Element-wise / Reduction}} \\
    \texttt{add}                 & 64M FP16                             & 90.8\% & 74.5\% \\
    \texttt{mul}                 & 64M FP16                             & 69.5\%  & 67.1\% \\
    \texttt{relu}                & $16384\!\times\!16384$ FP16          & 95.7\% & 68.1\% \\
    \texttt{softmax}             & $(4096,32768)$ FP16                  & 117.2\%  & 0.9\% \\
    \texttt{log\_softmax}        & $(4096,32768)$ FP16                  & 45.3\%  & 3.6\% \\
    \texttt{argmax}              & $(8192,32768)$ FP16                  & 25.5\%  & 5.9\%  \\
    \texttt{max\_reduction}      & $(8192,32768)$ FP16                  & 109.6\%  & 67.2\%  \\
    \texttt{mean\_reduction}     & $(8192,32768)$ FP32                  & 92.7\%  & 97.4\% \\
    \texttt{matrix\_transpose}   & $16384\!\times\!16384$ FP16          & \textbf{342.8\%}$^{\|}$ & 67.4\% \\
    \texttt{embedding}           & $(131072,1024)$ idx FP16             & \textbf{447.9\%}$^{\P}$ & 57.9\% \\
    \texttt{index\_select}       & $(65536,2048)$ idx FP16              & 59.2\%  & 56.6\% \\
    \bottomrule
  \end{tabular}}
  \smallskip
  \begin{minipage}{\columnwidth}
    \footnotesize
    $^{\dagger}$~TileLang \texttt{linear\_activation} uses a fused gate-and-activate pass; the speedup is genuine.
    $^{\ddagger}$~Triton \texttt{attention} uses a FlashAttention-style tiled kernel against PyTorch's unfused eager path; the comparison is excluded from the main GEMM analysis (\Cref{sec:eval:gemm}).
    $^{\S}$~Triton \texttt{rms\_norm} is faster than PyTorch's unfused \texttt{F.rms\_norm} reference; this is a baseline-choice artifact (\Cref{tab:summary} footnote).
    $^{\P}$~Triton \texttt{embedding} is faster than PyTorch's scatter-based eager lookup; attributed to coalesced vs.\ scattered memory access, not a library maturity gap.
    $^{\|}$~Triton \texttt{matrix\_transpose} is faster than PyTorch's eager transpose-copy path; a baseline-path effect analogous to \texttt{embedding}, not a library maturity gap.
  \end{minipage}
\end{table}

\section{Reproducibility Details}
\label{app:repro}

\subsection{Kernel Suite Provenance}
\label{app:repro:kernels}

\paragraph{Selection criteria}
We narrow the candidates from TritonBench~\cite{li2025tritonbench} and the TileLang example repository~\cite{wang2025tilelang} using four criteria applied in order.
\emph{(i) Forward-pass only}: backward and gradient kernels are excluded, as they are governed by different access patterns and reduction structures.
\emph{(ii) Stable library baseline}: each kernel must admit a well-defined cuBLAS, cuDNN, or PyTorch eager reference that computes the same function, making library efficiency a meaningful quantity.
\emph{(iii) Five-category coverage}: we retain kernels that populate the five operator categories (GEMM, attention, convolution, normalization, element-wise/reduction) rather than over-sampling any single class.
\emph{(iv) Canonical operators}: kernels with sparse inputs or runtime-dependent output shapes are excluded for lacking a stable baseline.

\paragraph{Per-kernel provenance}
\Cref{tab:provenance} lists all 22 kernels.
Triton implementations are drawn from TritonBench except \texttt{layer\_norm}, which follows the TorchInductor reference implementation.
All TileLang implementations are our re-implementations of the same operators.
PyTorch implementations are cuBLAS, cuDNN, or PyTorch-eager baselines as described in \cref{app:repro:baselines}.

\begin{table}[h]
  \centering
  \caption{Per-kernel provenance for the 22-kernel ViperBench suite.}
  \label{tab:provenance}
  \resizebox{\columnwidth}{!}{%
  \begin{tabular}{llll}
    \toprule
    Kernel & Category & Triton Source & TileLang Source \\
    \midrule
    \texttt{matmul}             & GEMM          & TritonBench         & Custom \\
    \texttt{batched\_matmul}    & GEMM          & TritonBench         & Custom \\
    \texttt{linear\_activation} & GEMM          & TritonBench         & Custom \\
    \midrule
    \texttt{attention}          & Attention     & TritonBench         & Custom \\
    \midrule
    \texttt{conv2d}             & Convolution   & TritonBench         & Custom \\
    \midrule
    \texttt{layer\_norm}        & Normalization & TorchInductor ref.  & Custom \\
    \texttt{rms\_norm}          & Normalization & TritonBench         & Custom \\
    \midrule
    \texttt{add}                & EW / Red.     & TritonBench         & Custom \\
    \texttt{mul}                & EW / Red.     & TritonBench         & Custom \\
    \texttt{relu}               & EW / Red.     & TritonBench         & Custom \\
    \texttt{leaky\_relu}        & EW / Red.     & TritonBench         & Custom \\
    \texttt{softmax}            & EW / Red.     & TritonBench         & Custom \\
    \texttt{log\_softmax}       & EW / Red.     & TritonBench         & Custom \\
    \texttt{logsumexp}          & EW / Red.     & TritonBench         & Custom \\
    \texttt{swiglu}             & EW / Red.     & TritonBench         & Custom \\
    \texttt{argmax}             & EW / Red.     & TritonBench         & Custom \\
    \texttt{max\_reduction}     & EW / Red.     & TritonBench         & Custom \\
    \texttt{mean\_reduction}    & EW / Red.     & TritonBench         & Custom \\
    \texttt{cross\_entropy}     & EW / Red.     & TritonBench         & Custom \\
    \texttt{matrix\_transpose}  & EW / Red.     & TritonBench         & Custom \\
    \texttt{index\_select}      & EW / Red.     & TritonBench         & Custom \\
    \texttt{embedding}          & EW / Red.     & TritonBench         & Custom \\
    \bottomrule
  \end{tabular}}
  \smallskip
  \begin{minipage}{\columnwidth}
    \small ``Custom'' denotes our re-implementation of the operator following the same interface as
    the TritonBench version; ``TorchInductor ref.''\ denotes an implementation modelled on the
    TorchInductor generated reference for that op.
    EW / Red.\ = element-wise or reduction.
  \end{minipage}
\end{table}

\subsection{Baseline Specifications}
\label{app:repro:baselines}

GEMM baselines use cuBLAS via \texttt{torch.matmul}.
Convolution baselines use PyTorch \texttt{nn.Conv2d} with default NCHW memory layout under standard cuDNN algorithm selection.
Normalization baselines use \texttt{F.layer\_norm} and \texttt{F.rms\_norm}.
Element-wise and reduction baselines use standard PyTorch eager execution.
For attention, we use PyTorch scaled dot-product attention with the FlashAttention backend enabled through \texttt{enable\_flash\_sdp(True)}.

\paragraph{GEMM notation and exclusions}
Per-shape GEMM latencies appear in \Cref{tab:perkernel}; $16384^2$ denotes a square $16384\times16384$ FP16 GEMM and $64\times128^2$ a batched GEMM of batch 64 over $128\times128$ matrices.
FP32 TileLang GEMM is excluded: \texttt{T.gemm} silently lowers to TF32, making FP32 results a precision artifact rather than a logic error (\Cref{sec:meth:profiling}).
\Cref{tab:summary} reports the category medians; the $16384^2$ Triton gap to 59.7\% reflects an under-populated auto-tune search space (RC2), while TileLang reaches 78.1\% on that shape via an explicit schedule.

\subsection{Measurement Protocol}
\label{app:repro:meas}

\paragraph{End-to-end timing}
We measure host-synchronized GPU execution time using \texttt{time.perf\_counter()} bracketed by \texttt{torch.cuda.synchronize()} calls.
Each measurement uses 10 warm-up iterations followed by 100 timed iterations; we report the median.
Results reflect GPU-side execution only, excluding host-side dispatch overhead, and each workload runs in isolation.

\paragraph{Clock-lock settings}
On the A100-SXM4-40GB, we pin graphics and memory clocks to 1215\,MHz and 1215\,MHz respectively via \texttt{nvidia-smi --lock-gpu-clocks} and \texttt{--lock-memory-clocks}.
The nominal maximum is 1410\,MHz, but that setting power-caps under the 400\,W board limit and causes clock throttling; 1215\,MHz is stable and yields run-to-run relative standard deviation of 0.0--0.9\% across 100-iteration timing windows.
On the GH200, clocks are locked at 1320\,MHz (graphics) and 2619\,MHz (memory).

\paragraph{Nsight Compute counter definitions}
We collect the following seven counters per kernel; each is gathered from a single \texttt{ncu} execution and verified within 2\% across five repeats.

\begin{enumerate}
  \item \textbf{Global-load efficiency} --- fraction of global memory transactions that serve requested bytes (1.0 = perfectly coalesced).
  \item \textbf{Sectors per request} --- average number of L1/L2 cache sectors accessed per memory instruction; lower values indicate better coalescing.
  \item \textbf{Registers per thread} --- register file allocation per thread; high values increase occupancy pressure and can trigger spill.
  \item \textbf{Register spills} --- bytes spilled from the register file to thread-local memory (L1/L2/DRAM); non-zero values denote register-pressure-induced latency.
  \item \textbf{Long-scoreboard stall cycles} --- warp stall cycles waiting for L2 or DRAM data; the dominant stall for memory-bound kernels.
  \item \textbf{Barrier stall cycles} --- warp stall cycles at thread-block synchronization barriers (\texttt{\_\_syncthreads} / \texttt{bar.sync}).
  \item \textbf{L2 sector hit rate} --- fraction of L2 cache sector accesses that hit; low rates indicate DRAM pressure.
\end{enumerate}

\paragraph{KernelBench evaluator harness}
For the passes-but-slow demonstration (\cref{sec:eval:passesslow}), we invoke KernelBench's \texttt{eval\_kernel\_against\_ref} function, which (i) overrides the candidate kernel's \texttt{get\_inputs} and \texttt{get\_init\_inputs} with those of the reference implementation so input shapes cannot be altered by the candidate, (ii) checks output equivalence on randomized inputs at the same per-dtype tolerances as our correctness suite (\cref{sec:meth:profiling}), and (iii) emits a warning only if the candidate's throughput exceeds $10\times$ the reference---a speedup-only guard that admits any slowdown.
We record pass/fail and measure slowdown separately using our clock-locked timing harness.

\subsection{Full Hardware and Software Stack}
\label{app:repro:stack}

\paragraph{Hardware}

\begin{table}[h]
  \centering
  \caption{GPU hardware specifications for the two primary architectures.}
  \label{tab:hw:specs}
  \resizebox{\columnwidth}{!}{%
  \begin{tabular}{lll}
    \toprule
    Property & A100-SXM4-40GB & GH200 \\
    \midrule
    Architecture      & Ampere (\texttt{sm\_80}) & Grace Hopper (\texttt{sm\_90}) \\
    SMs               & 108                      & 132 \\
    Memory            & 40\,GB HBM2e             & 96\,GB HBM3 \\
    Peak mem.\ BW     & $\sim$1.5\,TB/s          & $\sim$4.0\,TB/s \\
    Peak FP16 TC      & $\sim$312\,TFLOP/s       & $\sim$989.5\,TFLOP/s \\
    L2 cache          & 40\,MB                   & 60\,MB \\
    TDP               & 400\,W                   & 900\,W (SoC) \\
    Clock lock (g/m)  & 1215 / 1215\,MHz         & 1320 / 2619\,MHz \\
    NVIDIA Driver     & 610.43.02                & n/r\rlap{$^{*}$} \\
    CUDA Toolkit      & 12.8                     & 12.8 \\
    \bottomrule
  \end{tabular}}
  {\footnotesize $^{*}$~The GH200 node's driver version was not exported by the measurement harness logs; the CUDA runtime and all library versions are pinned by the artifact's lockfiles.}
\end{table}

\paragraph{Software stack}

\begin{table}[h]
  \centering
  \caption{Pinned software versions for all experiments.}
  \label{tab:sw:stack}
  \begin{tabular}{ll}
    \toprule
    Package & Version \\
    \midrule
    PyTorch        & 2.8.0+cu128 \\
    Triton         & 3.4.0 \\
    TileLang       & 0.1.6.post1 \\
    cuDNN          & bundled with PyTorch \\
    Nsight Compute & 2026.2.0 \\
    \bottomrule
  \end{tabular}
\end{table}

All experiments use default cuDNN algorithm selection flags and pre-warm the Triton auto-tuner before measurement.
A \texttt{requirements.txt} pinning this framework stack is distributed with the artifact; the pinned PyTorch wheel bundles its own CUDA runtime and cuDNN.

\subsection{Optimization Trajectory: TileLang LayerNorm}
\label{app:repro:trajectory}

To illustrate the correctness-gated iterative protocol of \cref{sec:meth:optimization}, \cref{tab:layernorm:trajectory} lists the full 18-iteration search for the TileLang LayerNorm kernel of \cref{sec:mitig:norm}.
Two edits dominate: iteration~1 replaces the manual \texttt{T.serial} reduction with a single \texttt{T.reduce} (RC0), and iteration~14 switches to native bfloat16 I/O, removing the intermediate \texttt{.float()} casts.
The intervening iterations confirm that thread count, tiling, and reduction algebra are second-order once the kernel is bandwidth-bound, and two configurations fail to compile---evidence that the search is genuine rather than a curated path.
Runtimes are on the separate development GPU (\cref{sec:mitigation}); the selected final kernel is re-timed on the A100-SXM4 in \cref{tab:mitigation}.

\begin{table}[t]
  \centering
  \caption{\textbf{TileLang LayerNorm optimization trajectory (development GPU).}
    Full 18-iteration correctness-gated search for the bfloat16 LayerNorm kernel of \cref{sec:mitig:norm}; every listed configuration is numerically correct except the two compile failures (iter 3, 7, shown ``---'').
    $E_\text{lib}$ is the latency ratio to PyTorch \texttt{F.layer\_norm} on the same GPU.
    \textbf{Bold} marks the two dominant edits: \texttt{T.serial}$\to$\texttt{T.reduce} (iter 1) and native bfloat16 I/O (iter 14).
    $^\dagger$ selected final configuration; re-timed on the A100-SXM4 as $0.270$~ms / $124\%$ in \cref{tab:mitigation}.}
  \label{tab:layernorm:trajectory}
  \footnotesize
  \setlength{\tabcolsep}{4pt}
  \begin{tabular}{@{}clrr@{}}
    \toprule
    Iter & Change & Runtime (ms) & $E_\text{lib}$ \\
    \midrule
    base & \texttt{T.serial} reduction & 1090 & 0.08\% \\
    \textbf{1} & \textbf{Replace \texttt{T.serial} with \texttt{T.reduce}} & \textbf{5.24} & \textbf{17.1\%} \\
    2 & Register copy, 256 threads & 5.22 & 17.2\% \\
    3 & 2D \texttt{block\_M}${=}4$, \texttt{reduce\_sum} dim 1 & --- & --- \\
    4 & 2D \texttt{T.copy} fix, \texttt{reduce\_sum} dim 1 & 5.22 & 17.2\% \\
    5 & 512 threads & 5.22 & 17.2\% \\
    6 & fp16 I/O, fp32 accumulate & 3.02 & 29.5\% \\
    7 & 2D merged $E[x^2]{-}E[x]^2$ & --- & --- \\
    8 & 1024 threads, fp16 I/O & 3.02 & 29.6\% \\
    9 & Tiled reduction \texttt{block\_N}${=}1024$ & 2.99 & 29.9\% \\
    10 & Two-pass tiled (sum${+}$sq in pass 1) & 3.00 & 29.8\% \\
    11 & $E[x^2]{-}E[x]^2$, single load & 3.02 & 29.6\% \\
    12 & Disable warp-specialized \texttt{pass\_configs} & 3.02 & 29.6\% \\
    13 & Store $x{-}\mu$ in fp16 to save registers & 3.02 & 29.6\% \\
    \textbf{14} & \textbf{Native bfloat16 I/O, no cast} & \textbf{0.893} & \textbf{99.8\%} \\
    15$^\dagger$ & \texttt{out\_idx} output allocation & 0.891 & 100.0\% \\
    16 & Full fp32 \texttt{row\_copy} (more regs) & 0.905 & 98.5\% \\
    17 & Restore iter-15 configuration & 0.893 & 99.8\% \\
    18 & 128 threads & 0.895 & 99.6\% \\
    \bottomrule
  \end{tabular}
\end{table}

%% file: references.bib
@inproceedings{tillet2019triton,
author = {Tillet, Philippe and Kung, H. T. and Cox, David},
title = {Triton: an intermediate language and compiler for tiled neural network computations},
year = {2019},
isbn = {9781450367196},
publisher = {Association for Computing Machinery},
address = {New York, NY, USA},
url = {https://doi.org/10.1145/3315508.3329973},
doi = {10.1145/3315508.3329973},
booktitle = {Proceedings of the 3rd ACM SIGPLAN International Workshop on Machine Learning and Programming Languages},
pages = {10–19},
numpages = {10},
location = {Phoenix, AZ, USA},
series = {MAPL 2019}
}

@misc{openai2021triton,
  title        = {Introducing {Triton}: Open-Source {GPU} Programming for Neural Networks},
  author       = {{OpenAI}},
  year         = {2021},
  howpublished = {\url{https://openai.com/index/triton/}}
}

@inproceedings{wang2025tilelang,
title={TileLang: Bridge Programmability and Performance in Modern Neural Kernels},
author={Lei Wang and Yu Cheng and Yining Shi and Zhiwen Mo and Zhengju Tang and Wenhao Xie and Tong Wu and Lingxiao Ma and Yuqing Xia and Jilong Xue and Fan Yang and Zhi Yang},
booktitle={The Fourteenth International Conference on Learning Representations},
year={2026},
url={https://openreview.net/forum?id=Jb1WkNSfUB}
}

@inproceedings{li2025tritonbench,
    title = "{T}riton{B}ench: Benchmarking Large Language Model Capabilities for Generating Triton Operators",
    author = "Li, Jianling  and
      Li, Shangzhan  and
      Gao, Zhenye  and
      Shi, Qi  and
      Li, Yuxuan  and
      Wang, Zefan  and
      Huang, Jiacheng  and
      Wang, Haojie  and
      Wang, Jianrong  and
      Han, Xu  and
      Liu, Zhiyuan  and
      Sun, Maosong",
    editor = "Che, Wanxiang  and
      Nabende, Joyce  and
      Shutova, Ekaterina  and
      Pilehvar, Mohammad Taher",
    booktitle = "Findings of the Association for Computational Linguistics: ACL 2025",
    month = jul,
    year = "2025",
    address = "Vienna, Austria",
    publisher = "Association for Computational Linguistics",
    url = "https://aclanthology.org/2025.findings-acl.1183/",
    doi = "10.18653/v1/2025.findings-acl.1183",
    pages = "23053--23066",
    ISBN = "979-8-89176-256-5"
}

@article{chetlur2014cudnn,
      title={cuDNN: Efficient Primitives for Deep Learning}, 
      author={Sharan Chetlur and Cliff Woolley and Philippe Vandermersch and Jonathan Cohen and John Tran and Bryan Catanzaro and Evan Shelhamer},
      year={2014},
      eprint={1410.0759},
      archivePrefix={arXiv},
      primaryClass={cs.NE},
      url={https://arxiv.org/abs/1410.0759}, 
}

@software{cublas2024,
  title        = {{cuBLAS}: Basic Linear Algebra on NVIDIA GPUs},
  author       = {{NVIDIA Corporation}},
  year         = {2026},
  howpublished = {\url{https://docs.nvidia.com/cuda/cublas/}}
}

@inproceedings{ragan-kelley2013halide,
author = {Ragan-Kelley, Jonathan and Barnes, Connelly and Adams, Andrew and Paris, Sylvain and Durand, Fr\'{e}do and Amarasinghe, Saman},
title = {Halide: a language and compiler for optimizing parallelism, locality, and recomputation in image processing pipelines},
year = {2013},
issue_date = {June 2013},
publisher = {Association for Computing Machinery},
address = {New York, NY, USA},
volume = {48},
number = {6},
issn = {0362-1340},
url = {https://doi.org/10.1145/2499370.2462176},
doi = {10.1145/2499370.2462176},
journal = {SIGPLAN Not.},
month = jun,
pages = {519–530},
numpages = {12}
}

@inproceedings{chen2018tvm,
author = {Tianqi Chen and Thierry Moreau and Ziheng Jiang and Lianmin Zheng and Eddie Yan and Haichen Shen and Meghan Cowan and Leyuan Wang and Yuwei Hu and Luis Ceze and Carlos Guestrin and Arvind Krishnamurthy},
title = {{TVM}: An Automated {End-to-End} Optimizing Compiler for Deep Learning},
booktitle = {13th USENIX Symposium on Operating Systems Design and Implementation (OSDI 18)},
year = {2018},
isbn = {978-1-939133-08-3},
address = {Carlsbad, CA},
pages = {578--594},
url = {https://www.usenix.org/conference/osdi18/presentation/chen},
publisher = {USENIX Association},
month = oct
}

@article{optimization_survey,
author = {Hijma, Pieter and Heldens, Stijn and Sclocco, Alessio and van Werkhoven, Ben and Bal, Henri E.},
title = {Optimization Techniques for GPU Programming},
year = {2023},
issue_date = {November 2023},
publisher = {Association for Computing Machinery},
address = {New York, NY, USA},
volume = {55},
number = {11},
issn = {0360-0300},
url = {https://doi.org/10.1145/3570638},
doi = {10.1145/3570638},
journal = {ACM Comput. Surv.},
month = mar,
articleno = {239},
numpages = {81}
}

@INPROCEEDINGS{kernelfusion,
  author={Wang, Guibin and Lin, YiSong and Yi, Wei},
  booktitle={2010 IEEE/ACM Int'l Conference on Green Computing and Communications \& Int'l Conference on Cyber, Physical and Social Computing},
  title={Kernel Fusion: An Effective Method for Better Power Efficiency on Multithreaded GPU}, 
  year={2010},
  volume={},
  number={},
  pages={344-350},
  doi={10.1109/GreenCom-CPSCom.2010.102}}

@ARTICLE{acceleration_survey,
  author={Shuvo, Md. Maruf Hossain and Islam, Syed Kamrul and Cheng, Jianlin and Morshed, Bashir I.},
  journal={Proceedings of the IEEE}, 
  title={Efficient Acceleration of Deep Learning Inference on Resource-Constrained Edge Devices: A Review}, 
  year={2023},
  volume={111},
  number={1},
  pages={42-91},
  doi={10.1109/JPROC.2022.3226481}}

@inproceedings{zheng2020ansor,
author = {Lianmin Zheng and Chengfan Jia and Minmin Sun and Zhao Wu and Cody Hao Yu and Ameer Haj-Ali and Yida Wang and Jun Yang and Danyang Zhuo and Koushik Sen and Joseph E. Gonzalez and Ion Stoica},
title = {Ansor: Generating {High-Performance} Tensor Programs for Deep Learning},
booktitle = {14th USENIX Symposium on Operating Systems Design and Implementation (OSDI 20)},
year = {2020},
isbn = {978-1-939133-19-9},
pages = {863--879},
url = {https://www.usenix.org/conference/osdi20/presentation/zheng},
publisher = {USENIX Association},
month = nov
}

@inproceedings{bik2022mlir,
author = {Katel, Navdeep and Khandelwal, Vivek and Bondhugula, Uday},
title = {MLIR-based code generation for GPU tensor cores},
year = {2022},
isbn = {9781450391832},
publisher = {Association for Computing Machinery},
address = {New York, NY, USA},
url = {https://doi.org/10.1145/3497776.3517770},
doi = {10.1145/3497776.3517770},
booktitle = {Proceedings of the 31st ACM SIGPLAN International Conference on Compiler Construction},
pages = {117–128},
numpages = {12},
location = {Seoul, South Korea},
series = {CC 2022}
}

@inproceedings{spector2024thunderkittens,
title={ThunderKittens: Simple, Fast, and \${\textbackslash}textit\{Adorable\}\$ Kernels},
author={Benjamin Frederick Spector and Simran Arora and Aaryan Singhal and Arjun Parthasarathy and Daniel Y Fu and Christopher Re},
booktitle={The Thirteenth International Conference on Learning Representations},
year={2025},
url={https://openreview.net/forum?id=0fJfVOSUra}
}

@misc{li2025tritonforge,
      title={TritonForge: Profiling-Guided Framework for Automated Triton Kernel Optimization}, 
      author={Haonan Li and Keyu Man and Partha Kanuparthy and Hanning Chen and Wei Sun and Sreen Tallam and Chenguang Zhu and Kevin Zhu and Zhiyun Qian},
      year={2025},
      eprint={2512.09196},
      archivePrefix={arXiv},
      primaryClass={cs.SE},
      url={https://arxiv.org/abs/2512.09196}, 
}

@inproceedings{zhou2021implicit,
  author={Zhou, Yangjie and Yang, Mengtian and Guo, Cong and Leng, Jingwen and Liang, Yun and Chen, Quan and Guo, Minyi and Zhu, Yuhao},
  booktitle={2021 IEEE International Symposium on Workload Characterization (IISWC)}, 
  title={Characterizing and Demystifying the Implicit Convolution Algorithm on Commercial Matrix-Multiplication Accelerators}, 
  year={2021},
  volume={},
  number={},
  pages={214-225},
  doi={10.1109/IISWC53511.2021.00029}}

@misc{triton2022conv591,
  title        = {Example {Conv2D} in {Triton}},
  author       = {{Triton Community}},
  year         = {2022},
  howpublished = {GitHub Discussion \#591, \url{https://github.com/triton-lang/triton/discussions/591}},
  note         = {Accessed: 2026-03-25}
}

@inproceedings{dao2023flashattention2,
      title={FlashAttention-2: Faster Attention with Better Parallelism and Work Partitioning}, 
      author={Tri Dao},
      year={2023},
      eprint={2307.08691},
      archivePrefix={arXiv},
      primaryClass={cs.LG},
      url={https://arxiv.org/abs/2307.08691}, 
}

@inproceedings{li2023torchcompile,
author = {Ansel, Jason and Yang, Edward and He, Horace and Gimelshein, Natalia and Jain, Animesh and Voznesensky, Michael and Bao, Bin and Bell, Peter and Berard, David and Burovski, Evgeni and Chauhan, Geeta and Chourdia, Anjali and Constable, Will and Desmaison, Alban and DeVito, Zachary and Ellison, Elias and Feng, Will and Gong, Jiong and Gschwind, Michael and Hirsh, Brian and Huang, Sherlock and Kalambarkar, Kshiteej and Kirsch, Laurent and Lazos, Michael and Lezcano, Mario and Liang, Yanbo and Liang, Jason and Lu, Yinghai and Luk, C. K. and Maher, Bert and Pan, Yunjie and Puhrsch, Christian and Reso, Matthias and Saroufim, Mark and Siraichi, Marcos Yukio and Suk, Helen and Zhang, Shunting and Suo, Michael and Tillet, Phil and Zhao, Xu and Wang, Eikan and Zhou, Keren and Zou, Richard and Wang, Xiaodong and Mathews, Ajit and Wen, William and Chanan, Gregory and Wu, Peng and Chintala, Soumith},
title = {PyTorch 2: Faster Machine Learning Through Dynamic Python Bytecode Transformation and Graph Compilation},
year = {2024},
isbn = {9798400703850},
publisher = {Association for Computing Machinery},
address = {New York, NY, USA},
url = {https://doi.org/10.1145/3620665.3640366},
doi = {10.1145/3620665.3640366},
booktitle = {Proceedings of the 29th ACM International Conference on Architectural Support for Programming Languages and Operating Systems, Volume 2},
pages = {929–947},
numpages = {19},
location = {La Jolla, CA, USA},
series = {ASPLOS '24}
}

@misc{vasilache2022composable,
      title={Composable and Modular Code Generation in MLIR: A Structured and Retargetable Approach to Tensor Compiler Construction}, 
      author={Nicolas Vasilache and Oleksandr Zinenko and Aart J. C. Bik and Mahesh Ravishankar and Thomas Raoux and Alexander Belyaev and Matthias Springer and Tobias Gysi and Diego Caballero and Stephan Herhut and Stella Laurenzo and Albert Cohen},
      year={2022},
      eprint={2202.03293},
      archivePrefix={arXiv},
      primaryClass={cs.PL},
      url={https://arxiv.org/abs/2202.03293}, 
}

@inproceedings{wolfe1989more,
author = {Wolfe, M.},
title = {More iteration space tiling},
year = {1989},
isbn = {0897913418},
publisher = {Association for Computing Machinery},
address = {New York, NY, USA},
url = {https://doi.org/10.1145/76263.76337},
doi = {10.1145/76263.76337},
booktitle = {Proceedings of the 1989 ACM/IEEE Conference on Supercomputing},
pages = {655–664},
numpages = {10},
location = {Reno, Nevada, USA},
series = {Supercomputing '89}
}

@misc{nsightcompute2024,
  title        = {{NVIDIA} Nsight Compute},
  author       = {{NVIDIA Corporation}},
  year         = {2024},
  howpublished = {\url{https://developer.nvidia.com/nsight-compute}}
}

@software{cutlass2023,
  title        = {{CUTLASS}: CUDA Templates and Python DSLs for High-Performance Linear Algebra},
  author       = {{NVIDIA Corporation}},
  year         = {2017},
  howpublished = {\url{https://github.com/NVIDIA/cutlass}}
}

@INPROCEEDINGS{reddi2020mlperf,
  author={Reddi, Vijay Janapa and Cheng, Christine and Kanter, David and Mattson, Peter and Schmuelling, Guenther and Wu, Carole-Jean and Anderson, Brian and Breughe, Maximilien and Charlebois, Mark and Chou, William and Chukka, Ramesh and Coleman, Cody and Davis, Sam and Deng, Pan and Diamos, Greg and Duke, Jared and Fick, Dave and Gardner, J. Scott and Hubara, Itay and Idgunji, Sachin and Jablin, Thomas B. and Jiao, Jeff and John, Tom St. and Kanwar, Pankaj and Lee, David and Liao, Jeffery and Lokhmotov, Anton and Massa, Francisco and Meng, Peng and Micikevicius, Paulius and Osborne, Colin and Pekhimenko, Gennady and Rajan, Arun Tejusve Raghunath and Sequeira, Dilip and Sirasao, Ashish and Sun, Fei and Tang, Hanlin and Thomson, Michael and Wei, Frank and Wu, Ephrem and Xu, Lingjie and Yamada, Koichi and Yu, Bing and Yuan, George and Zhong, Aaron and Zhang, Peizhao and Zhou, Yuchen},
  booktitle={2020 ACM/IEEE 47th Annual International Symposium on Computer Architecture (ISCA)}, 
  title={MLPerf Inference Benchmark}, 
  year={2020},
  volume={},
  number={},
  pages={446-459},
  doi={10.1109/ISCA45697.2020.00045}}

@article{williams2009roofline,
author = {Williams, Samuel and Waterman, Andrew and Patterson, David},
title = {Roofline: an insightful visual performance model for multicore architectures},
year = {2009},
issue_date = {April 2009},
publisher = {Association for Computing Machinery},
address = {New York, NY, USA},
volume = {52},
number = {4},
issn = {0001-0782},
url = {https://doi.org/10.1145/1498765.1498785},
doi = {10.1145/1498765.1498785},
journal = {Commun. ACM},
month = apr,
pages = {65–76},
numpages = {12}
}

@INPROCEEDINGS{lattner2021mlir,
  author={Lattner, Chris and Amini, Mehdi and Bondhugula, Uday and Cohen, Albert and Davis, Andy and Pienaar, Jacques and Riddle, River and Shpeisman, Tatiana and Vasilache, Nicolas and Zinenko, Oleksandr},
  booktitle={2021 IEEE/ACM International Symposium on Code Generation and Optimization (CGO)}, 
  title={MLIR: Scaling Compiler Infrastructure for Domain Specific Computation}, 
  year={2021},
  volume={},
  number={},
  pages={2-14},
  doi={10.1109/CGO51591.2021.9370308}}

@misc{kernelbench,
      title={{KernelBench}: Can {LLMs} Write Efficient {GPU} Kernels?},
      author={Anne Ouyang and Simon Guo and Simran Arora and Alex L. Zhang and William Hu and Christopher R\'{e} and Azalia Mirhoseini},
      year={2025},
      eprint={2502.10517},
      archivePrefix={arXiv},
      primaryClass={cs.LG},
      url={https://arxiv.org/abs/2502.10517},
}

@misc{ako2026,
  title        = {{AKO}: Agentic Kernel Optimization},
  author       = {Shuxiao Xie and Shuyang Xie and Dezhi Ran and Wei Yang and Tao Xie},
  year         = {2026},
  howpublished = {\url{https://tongminglaic.github.io/AKO}},
  note         = {Technical report}
}

@misc{sakana2025aicuda,
  title        = {The {AI} {CUDA} Engineer: Agentic {CUDA} Kernel Discovery, Optimization and Composition},
  author       = {{Sakana AI}},
  year         = {2025},
  howpublished = {\url{https://sakana.ai/ai-cuda-engineer/}},
  note         = {The initially reported $150\times$ speedups were later traced to kernels exploiting the evaluation harness to bypass correctness checking; see the project's postmortem update}
}

@inproceedings{liu2024evalperf,
  title     = {Evaluating Language Models for Efficient Code Generation},
  author    = {Liu, Jiawei and Xie, Songrun and Wang, Junhao and Wei, Yuxiang and Ding, Yifeng and Zhang, Lingming},
  booktitle = {First Conference on Language Modeling (COLM)},
  year      = {2024},
  url       = {https://openreview.net/forum?id=IBCBMeAhmC}
}

@inproceedings{du2024mercury,
  title     = {Mercury: A Code Efficiency Benchmark for Code Large Language Models},
  author    = {Du, Mingzhe and Luu, Anh Tuan and Ji, Bin and Liu, Qian and Ng, See-Kiong},
  booktitle = {Advances in Neural Information Processing Systems (NeurIPS), Datasets and Benchmarks Track},
  year      = {2024}
}

@inproceedings{huang2024effibench,
  title     = {{EffiBench}: Benchmarking the Efficiency of Automatically Generated Code},
  author    = {Huang, Dong and Qing, Yuhao and Shang, Weiyi and Cui, Heming and Zhang, Jie M.},
  booktitle = {Proceedings of the 38th International Conference on Neural Information Processing Systems},
  series    = {NIPS '24},
  articleno = {367},
  numpages  = {39},
  year      = {2024},
  isbn      = {9798331314385},
  publisher = {Curran Associates Inc.},
  address   = {Red Hook, NY, USA},
  location  = {Vancouver, BC, Canada}
}

@misc{qiu2024enamel,
  title         = {How Efficient is {LLM}-Generated Code? A Rigorous \& High-Standard Benchmark},
  author        = {Qiu, Ruizhong and Zeng, Weiliang Will and Ezick, James and Lott, Christopher and Tong, Hanghang},
  year          = {2025},
  eprint        = {2406.06647},
  archivePrefix = {arXiv},
  primaryClass  = {cs.SE},
  url           = {https://arxiv.org/abs/2406.06647}
}

@inproceedings{jin2012understanding,
  title     = {Understanding and Detecting Real-World Performance Bugs},
  author    = {Jin, Guoliang and Song, Linhai and Shi, Xiaoming and Scherpelz, Joel and Lu, Shan},
  booktitle = {Proceedings of the 33rd ACM SIGPLAN Conference on Programming Language Design and Implementation (PLDI)},
  pages     = {77--88},
  year      = {2012},
  doi       = {10.1145/2254064.2254075}
}

@inproceedings{nistor2013discovering,
  title     = {Discovering, Reporting, and Fixing Performance Bugs},
  author    = {Nistor, Adrian and Jiang, Tian and Tan, Lin},
  booktitle = {Proceedings of the 10th Working Conference on Mining Software Repositories (MSR)},
  pages     = {237--246},
  year      = {2013},
  doi       = {10.1109/MSR.2013.6624035}
}

@misc{rathnasuriya2026tilebugs,
  title         = {Characterizing Real-World Bugs in Tile Programs for Automated Bug Detection},
  author        = {Rathnasuriya, Ravishka and Song, Zihe and Majoju, Nidhi and Moharir, Aaryaa and Li, Tingxi and Yang, Wei and Xie, Tao},
  year          = {2026},
  eprint        = {2605.19652},
  archivePrefix = {arXiv},
  primaryClass  = {cs.SE},
  url           = {https://arxiv.org/abs/2605.19652}
}

@article{barr2015oracle,
  title   = {The Oracle Problem in Software Testing: A Survey},
  author  = {Barr, Earl T. and Harman, Mark and McMinn, Phil and Shahbaz, Muzammil and Yoo, Shin},
  journal = {IEEE Transactions on Software Engineering},
  volume  = {41},
  number  = {5},
  pages   = {507--525},
  year    = {2015},
  doi     = {10.1109/TSE.2014.2372785}
}
